\documentstyle[11pt,titlepage,times]{article}
\mathcode`\0="0030      %
\mathcode`\1="0031
\mathcode`\2="0032
\mathcode`\3="0033
\mathcode`\4="0034
\mathcode`\5="0035
\mathcode`\6="0036
\mathcode`\7="0037
\mathcode`\8="0038
\mathcode`\9="0039

\makeatletter
\def\@begintheorem#1#2{\trivlist\item[\hskip\labelsep{\bf #1\ #2}]}
\def\foobarpt{\textfont\z@\tenrm 
  \scriptfont\z@\ninrm \scriptscriptfont\z@\sevrm
\textfont\@ne\tenmi \scriptfont\@ne\ninmi \scriptscriptfont\@ne\sevmi
\textfont\tw@\tensy \scriptfont\tw@\ninsy \scriptscriptfont\tw@\sevsy
\textfont\thr@@\tenex \scriptfont\thr@@\tenex \scriptscriptfont\thr@@\tenex
\def\unboldmath{\everymath{}\everydisplay{}\@nomath\unboldmath
          \textfont\@ne\tenmi 
          \textfont\tw@\tensy \textfont\lyfam\tenly
          \@boldfalse}\@boldfalse
\def\boldmath{\@ifundefined{tenmib}{\global\font\tenmib\@mbi\@magscale1\global
        \font\tensyb\@mbsy \@magscale1\global\font
         \tenlyb\@lasyb\@magscale1\relax\@addfontinfo\@xiipt
              {\def\boldmath{\everymath
                {\mit}\everydisplay{\mit}\@prtct\@nomathbold
                \textfont\@ne\tenmib \textfont\tw@\tensyb 
                \textfont\lyfam\tenlyb\@prtct\@boldtrue}}}{}\@xiipt\boldmath}%
\def\prm{\fam\z@\tenrm}%
\def\pit{\fam\itfam\tenit}\textfont\itfam\tenit \scriptfont\itfam\ninit
   \scriptscriptfont\itfam\sevit
\def\psl{\fam\slfam\tensl}\textfont\slfam\tensl 
     \scriptfont\slfam\tensl \scriptscriptfont\slfam\tensl
\def\pbf{\fam\bffam\tenbf}\textfont\bffam\tenbf 
   \scriptfont\bffam\ninbf \scriptscriptfont\bffam\ninbf 
\def\ptt{\fam\ttfam\tentt}\textfont\ttfam\tentt
   \scriptfont\ttfam\nintt \scriptscriptfont\ttfam\nintt 
\def\psf{\fam\sffam\tensf}\textfont\sffam\tensf
    \scriptfont\sffam\tensf \scriptscriptfont\sffam\tensf
\def\psc{\@getfont\psc\scfam\@xiipt{\@mcsc\@magscale1}}%
\def\ly{\fam\lyfam\tenly}\textfont\lyfam\tenly 
   \scriptfont\lyfam\ninly \scriptscriptfont\lyfam\sevly
 \@setstrut \rm}

\newcommand{\singlespacing}{\let\CS=
\@currsize\renewcommand{\baselinestretch}{1}\tiny\CS}
\newcommand{\singlespacingplus}{\let\CS=
\@currsize\renewcommand{\baselinestretch}{1.25}\tiny\CS}
\newcommand{\doublespacing}{\let\CS=
\@currsize\renewcommand{\baselinestretch}{1.75}\tiny\CS}
\newcommand{\draftspacing}{\let\CS=
\@currsize\renewcommand{\baselinestretch}{2.0}\tiny\CS}

\newcommand{\niceonespacing}{\let\CS=\@currsize\renewcommand{\baselinestretch}{1.1}\tiny\CS}\newcommand{\nicetwospacing}{\let\CS=\@currsize\renewcommand{\baselinestretch}{1.2}\tiny\CS}
\newcommand{\nicethreespacing}{\let\CS=\@currsize\renewcommand{\baselinestretch}{1.3}\tiny\CS}
\newcommand{\singlespacingplusplus}{\let\CS=\@currsize\renewcommand{\baselinestretch}{1.35}\tiny\CS}
\newcommand{\nicefivespacing}{\let\CS=\@currsize\renewcommand{\baselinestretch}{1.5}\tiny\CS}
\newcommand{\nicesixpacing}{\let\CS=\@currsize\renewcommand{\baselinestretch}{1.6}\tiny\CS}

\makeatother

\makeatletter
\def\@cite#1#2{[#1\if@tempswa , #2\fi]}
\makeatother

\newcommand\seq{\subseteq}
\renewcommand\.{\cdot}
\newcommand\<{\langle}
\renewcommand\>{\rangle}

\newcommand\Lora{\, \Longrightarrow \ }
\newcommand\Lolra{\ \Longleftrightarrow \ }

\newcommand\condition{\, | \,}
\newcommand{\sigmastar}{\mbox{$\Sigma^\ast$}}
\newcommand{\equalsdef}{\stackrel{\mbox{\protect\scriptsize df}}{=}}

\newcommand\tweak{\hspace*{1pt}}

\newcommand\N{{\rm I\!N}}

\newcommand\p{\mbox{\rm P}}
\newcommand\pinf{\mbox{\rm P}_{\infty}}
\newcommand\fp{\mbox{\rm FP}}
\newcommand\np{\mbox{\rm NP}}

\newcommand\conp{\mbox{\rm coNP}}

\newcommand\co{\mbox{\rm co}}

\newcommand\x{{\sf SP}}

\newcommand\easyforallio{\mbox{EASY}^{\forall}_{\mbox{\protect\scriptsize io}}}
\newcommand\easyforall{\mbox{EASY}^{\forall}_{\forall}}
\newcommand\easyforallae{\mbox{EASY}^{\forall}_{\mbox{\protect\scriptsize ae}}}
\newcommand\easyexistsio{\mbox{EASY}^{\exists}_{\mbox{\protect\scriptsize io}}}
\newcommand\easyexists{\mbox{EASY}^{\exists}_{\forall}}
\newcommand\easyexistsae{\mbox{EASY}^{\exists}_{\mbox{\protect\scriptsize ae}}}

\newcommand\existsio{\exists^{\mbox{\protect\scriptsize io}}}
\newcommand\forallae{\forall^{\mbox{\protect\scriptsize ae}}}

\newtheorem{theorem}{Theorem}[section]
\newtheorem{corollary}[theorem]{Corollary}

\newtheorem{observation}[theorem]{Observation}

\newtheorem{proposition}[theorem]{Proposition}
\newtheorem{definition}[theorem]{Definition}

\newtheorem{remark}[theorem]{Remark}
\newtheorem{claim}[theorem]{Claim}

\newenvironment{block}{\begin{list}{\hbox{}}{\leftmargin 1em
    \itemindent -1em \topsep 0pt \itemsep 0pt \partopsep 0pt}}{\end{list}}

\load{\scriptsize}{\cal}
\load{\tiny}{\cal}

\begin{document}

\bibliographystyle{alpha}

\title{Easy Sets and Hard Certificate Schemes}

\author{ 
{\em  Lane A. Hemaspaandra\/}\thanks{Supported 
in part by 
grants NSF-INT-9513368/DAAD-315-PRO-fo-ab, 
NSF-CCR-8957604, 
NSF-INT-9116781/\allowbreak{}JSPS-ENGR-207, and
NSF-CCR-9322513.  
Work done in part while visiting the Friedrich-Schiller-Universit\"at Jena.
} \\
Department of Computer Science \\
University of Rochester \\
Rochester, NY 14627 \\
lane@cs.rochester.edu
\and 
{\em  J\"org Rothe\/}\thanks{Supported in part 
by grant
NSF-INT-9513368/DAAD-315-PRO-fo-ab.
Work done in part while visiting the University of Rochester
and Le Moyne College, Syracuse.
} \\
Institut f\"ur Informatik \\
Friedrich-Schiller-Universit\"at Jena \\
07743 Jena, Germany \\
rothe@informatik.uni-jena.de
\and
{\em  Gerd Wechsung\/}\thanks{Supported in part 
by grant
NSF-INT-9513368/DAAD-315-PRO-fo-ab.
Work done in part while visiting the University of Rochester
and Le Moyne College, Syracuse.
} \\
Institut f\"ur Informatik \\
Friedrich-Schiller-Universit\"at Jena \\
07743 Jena, Germany \\
wechsung@minet.uni-jena.de
}

\newcount\hour  \newcount\minutes  \hour=\time  \divide\hour by 60
\minutes=\hour  \multiply\minutes by -60  \advance\minutes by \time
\def\mmmddyyyy{\ifcase\month\or Jan\or Feb\or Mar\or Apr\or May\or Jun\or Jul\or
  Aug\or Sep\or Oct\or Nov\or Dec\fi \space\number\day, \number\year}
\def\hhmm{\ifnum\hour<10 0\fi\number\hour :%
  \ifnum\minutes<10 0\fi\number\minutes}
\def\Draft{{\it Draft of \mmmddyyyy}}

\date{}

\typeout{WARNING:  BADNESS used to suppress reporting!  Beware!!}
\hbadness=3000%
\vbadness=10000 %

\setcounter{footnote}{0}
{\singlespacing\maketitle}

\begin{center}
{\large\bf Abstract}
\end{center}
\begin{quotation}
{\singlespacing

\noindent
Can easy sets only have easy certificate schemes?  In this paper, we
study the class of sets that, for all NP certificate schemes (i.e., NP
machines), always have easy acceptance certificates (i.e., accepting
paths) that can be computed in polynomial time.  We also study the
class of sets that, for all NP certificate schemes, infinitely often
have easy acceptance certificates.  

In particular, we provide equivalent characterizations of these
classes in terms of relative generalized Kolmogorov complexity,
showing that they are robust.  We also provide structural
conditions---regarding immunity and class collapses---that put upper
and lower bounds on the sizes of these two classes.  Finally, we
provide negative results showing that some of our positive claims are
optimal with regard to being relativizable.  Our negative results are
proven using a novel observation: we show that the classical
``wide spacing'' oracle construction technique yields instant
non-bi-immunity results.  Furthermore, we establish a result that
improves upon Baker, Gill, and Solovay's classical result that
\mbox{$\np \neq \p = \np \cap \conp$} holds in some relativized world.

}
\end{quotation}

\nicetwospacing
\setcounter{page}{1}
\pagestyle{plain}

\sloppy

\section{Introduction}

Borodin and Demers~\cite{bor-dem:t:selfreducibility}
proved the following result.
\begin{theorem} \label{thm:bor-dem} {\cite{bor-dem:t:selfreducibility}} \quad
If $\np \cap \conp \neq \p$, then there exists a set $L$
such that
\begin{enumerate}
\item $L \in \p$,
\item $L \seq \mbox{SAT}$, and
\item For no polynomial-time computable function $f$ does it 
hold that: for each $F \in L$, $f(F)$ outputs a satisfying
assignment of $F$.  
\end{enumerate}
\end{theorem}

That is, under a hypothesis most complexity 
theoreticians would guess to be true,
it follows that there is a set of satisfiable formulas for which it is
trivial to determine they are satisfiable, yet it is hard to determine
why (i.e., via what satisfying assignment) they are satisfiable.

Motivated by their work, this paper seeks to study,
complexity-theoretically, the classes of sets that do or do not have
easy certificates.  In particular, we are interested in 
the following four classes.
$\easyforall$ 
is the class of sets $L$ such that for each NP machine~$M$ accepting them,
there is a polynomial-time computable function~$f_M$ such that 
for each~$x\in L$, $f_M(x)$ outputs an accepting path 
of~$M(x)$.  That is, $\easyforall$ is the class of sets that
for all certificate schemes, have easy certificates for all 
elements of the set.  We can analogously define
$\easyforallio$,
$\easyexists$, and 
$\easyexistsio$.
However, we note that 
$\easyexists = \p$ and 
$\easyexistsio$ equals the class of non-P-immune NP sets.
Regarding the two $\mbox{EASY}^{\forall}_{...}$ classes,
we provide equivalent characterizations 
of the classes in terms of relative generalized Kolmogorov complexity, 
showing that they are robust.  We also provide
structural conditions---regarding immunity and class
collapses---that put upper and lower bounds on the sizes of these
two classes.  Finally, we provide negative results showing
that some of our positive claims are optimal with regard
to being relativizable.  Our negative results are proven using
a novel observation:  we show that the classical ``wide spacing''
oracle 
construction technique 
yields instant non-bi-immunity results.
Furthermore, we establish a result that improves upon Baker, Gill, and 
Solovay's classical result that \mbox{$\np \neq \p = \np
  \cap \conp$} holds in some
relativized world~\cite{bak-gil-sol:j:p-np-oracle}, 
and that in addition links their result with 
the above-stated result of Borodin and Demers.

\section{Definitions and Robustness}

For the standard notations and the complexity-theoretical concepts used 
in this paper we refer to some standard text book on computational 
complexity such as~\cite{hop-ull:b:automata,bal-dia-gab:b:sct,bov-cre:b:complexity,pap:b:complexity}.  
Fix the alphabet $\Sigma =
\{0,1\}$. $\sigmastar$ is the set of all strings over $\Sigma$.  For
each string $u\in \Sigma^{*}$, $|u|$ denotes the length of~$u$.  The
empty string is denoted by $\epsilon$.  As is standard, the notation
$\existsio x$ (respectively, $\forallae x$) means ``there exist
infinitely many $x$'' (``for all but finitely many $x$'').  For each
set $L\subseteq \Sigma^{*}$, $\|L\|$ denotes the cardinality of $L$
and $\overline{L} = \Sigma^{*}-L$ denotes the complement of $L$\@.
For any class $\tweak\cal C$ of sets, define $\co\tweak{\cal C}
\equalsdef \{L\,|\,\overline{L}\in \tweak{\cal C}\}$.  $L^{=n}$
($L^{\leq n}$) is the set of all strings in $L$ having length~$n$
(less than or equal to~$n$).  Let $\Sigma^n$ and $\Sigma^{\leq n}$ be
shorthands for $(\sigmastar)^{=n}$ and $(\sigmastar)^{\leq n}$,
respectively.  Let FINITE be the class of all finite sets.  To encode
a pair of strings, we use a polynomial-time computable, one-one, onto
pairing function, $\<\.,\.\>:\sigmastar\times \sigmastar
\,\rightarrow\, \sigmastar$, that has polynomial-time computable
inverses. 
Denoting the set of non-negative integers by~$\N$ and
using the standard correspondence between $\Sigma^*$ and $\N$, we will
view $\<\.,\.\>$ also as a pairing function mapping $\N \times \N$
onto $\N$\@.  Let $\leq_{\mbox{\protect\scriptsize lex}}$ denote the
standard quasi-lexicographical ordering on $\Sigma^{*}$, that is, for
strings $x$ and $y$, $x\leq_{\mbox{\protect\scriptsize lex}} y$ if
either $x=y$, or $|x|<|y|$, or $(|x|=|y|$ and there exists some $z\in
\Sigma^{*}$ such that $x=z0u$ and $y=z1v)$.  $x
<_{\mbox{\protect\scriptsize lex}} y$ indicates that
$x\leq_{\mbox{\protect\scriptsize lex}} y$ but $x\neq y$.

We will abbreviate ``polynomial-time deterministic (nondeterministic)
Turing machine'' by DPM (NPM). For any Turing machine $M$, $L(M)$
denotes the set of strings accepted by $M$, and the notation $M(x)$
means ``$M$ on input~$x$.''  For any oracle Turing machine~$M$ and any
oracle set~$A$, $L(M^A)$ denotes the set of strings accepted by~$M$
relative to~$A$, and the notation $M^A(x)$ means ``$M^A$ on
input~$x$.'' For any NPM~$N$ and any input~$x$, we assume that all
paths of $N(x)$ are suitably encoded by strings over $\Sigma$\@. An
NPM~$N$ is said to be {\em normalized\/} if there exists a
polynomial~$q$ such that for all $n$, $q(n) \geq n$ and, on each input
of length~$n$, all paths of length $q(n)$ exist in the computation 
of~$N(x)$, and $N(x)$ has only paths of length~$q(n)$. Unless otherwise
stated, all NPMs considered in this paper are required to be
normalized. For any NPM~$N$ and any input~$x$, we denote the set of
accepting paths of $N(x)$ by $\mbox{acc}_{N}(x)$.  

P (respectively,~NP) is the class of all sets that are accepted by
some DPM (NPM).  Let $\pinf$ denote the class $\p - \mbox{FINITE}$ of
all infinite P sets.  FP denotes the class of all polynomial-time
computable functions. For any complexity class $\tweak\cal C$, a set
$L$ is said to be $\tweak\cal C${\em -immune\/} if $L$ is infinite and
no infinite subset of $L$ is in $\tweak\cal C$. Let $\tweak\cal
C$-immune denote the class of all $\tweak\cal C$-immune sets.  A set
$L$ is said to be $\tweak\cal C${\em -bi-immune\/} if both $L$ and
$\overline{L}$ are $\tweak\cal C$-immune. Let $\tweak\cal C$-bi-immune
denote the class of all $\tweak\cal C$-bi-immune sets.  For classes
$\tweak\cal C$ and $\tweak\cal D$ of sets, $\tweak\cal D$ is said to
be $\tweak\cal C$-immune (respectively, $\tweak\cal C$-bi-immune) if
$\tweak{\cal D} \cap ({\cal C}\mbox{-immune}) \neq \emptyset$
(respectively, if $\tweak{\cal D} \cap ({\cal C}\mbox{-bi-immune}) \neq
\emptyset$).

As a notational convention, for any NPM $N$, we will say ``$N$ has
always (respectively, $N$ has infinitely often) easy certificates'' to
mean that (the encoding of) an accepting path of $N(x)$ can be printed
in polynomial time for each string $x \in L(N)$ (respectively, for
infinitely many $x \in L(N)$). Similarly, we will say ``$N$ has only
(respectively, $N$ has infinitely often) hard certificates'' to mean
that no FP function is able to output (the encoding of) an accepting
path of $N(x)$ for each string \mbox{$x \in L(N)$} (respectively, for
infinitely many \mbox{$x \in L(N)$}). This is more formally stated in
Definition~\ref{def:easy} below that introduces the classes
$\easyforall$, $\easyforallio$, $\easyexists$, and $\easyexistsio$ of
sets for which all (or some) NP certificate schemes accepting the set
have always (or infinitely often) easy certificates.

\begin{definition}
\label{def:easy}
\quad
Let $L \seq \sigmastar$ be a set.
\begin{enumerate}
\item \label{def:easy:1} $L \in \easyforall$ if and only if 
\begin{enumerate}
\item $L \in \np$, and

\item $(\forall N)\, [ (\mbox{$N$ is NPM with } L(N) = L)
  \Longrightarrow\ (\exists f_{N} \in \fp)\, (\forall x \in L)\, [
  f_{N}(x) \in \mbox{acc}_{N}(x)]]$.
\end{enumerate}

\item \label{def:easy:2} $L \in \easyforallio$ if and only if either $L$
  is finite, or
\begin{enumerate}
\item $L \in \np$, and

\item $(\forall N)\, [ (\mbox{$N$ is NPM with } L(N) = L)
  \Longrightarrow\ (\exists f_{N} \in \fp)\, (\existsio x \in L)\, [
  f_{N}(x) \in \mbox{acc}_{N}(x)]]$.
\end{enumerate}

\item \label{def:easy:3} $L \in \easyexists$ if and only if 
\begin{enumerate}
\item $L \in \np$, and

\item $(\exists \mbox{NPM $N$})\, [ L(N) = L \,\wedge\, (\exists f_{N}
  \in \fp)\, (\forall x \in L)\, [ f_{N}(x) \in \mbox{acc}_{N}(x)]]$.
\end{enumerate}

\item \label{def:easy:4} $L \in \easyexistsio$ if and only if either $L$
  is finite, or
\begin{enumerate}
\item $L \in \np$, and

\item $(\exists \mbox{NPM $N$})\, [ L(N) = L \,\wedge\, (\exists f_{N}
  \in \fp)\, (\existsio x \in L)\, [ f_{N}(x) \in
  \mbox{acc}_{N}(x)]]$.
\end{enumerate}
\end{enumerate}
\end{definition}

\begin{remark}
\begin{enumerate}
\item It is easy to see that both $\easyforall$ and $\easyexists$
  contain all finite sets.  On the other hand, the
  $\mbox{EASY}^{...}_{\mbox{\scriptsize io}}$ classes are {\em
    defined\/} so as to also contain all finite sets; this is just for
  uniformity and since we feel that it is reasonable to require that
  the finite sets satisfy any suggested notion of ``easy sets.''
  Note that the $\mbox{EASY}^{...}_{\mbox{\scriptsize io}}$ classes
  capture the idea that sets having {\em correct\/} NP programs {\em
    for all inputs\/} may have {\em easy to compute\/} NP certificates
  just on {\em infinitely many\/} inputs, while possibly having only
  hard to compute NP certificates on infinitely many other inputs. Let
  us give some more motivation for these two classes.  In particular,
  we are interested in comparing the ``io'' notions with the
  corresponding ``$\forall$'' notions. While $\easyexistsio$ turns out
  to be a class that is reasonably characterizable in terms of immunity
  (cf.~Theorem~\ref{thm:inclusions}.\ref{thm:existsio} below),
  $\easyforallio$ plays an important role for intermediate conditions
  between certain immunity and other statements belonging to the
  implications that will be proven as the main results of the next
  section and that are summarized in Figure~\ref{fig:arrows}.

\item Note also that we can analogously define $\easyforallae$ and
  $\easyexistsae$ by using the quantification ``$\forallae x \in L$''
  rather than ``$\forall x \in L$'' in Parts~\ref{def:easy:1}
  and~\ref{def:easy:3} of the above definition.  However, since the
  classes $\easyforall$ and $\easyexists$ (as are most complexity
  classes) are closed under finite variations, it is clear that
  \mbox{$\easyforallae = \easyforall$} and $\easyexistsae =
  \easyexists$.  Moreover, we show below that $\easyexists = \p$ and
  that $\easyexistsio$ equals the class of all non-P-immune NP sets,
  and we therefore will not further discuss these two classes in this
  paper.
  \end{enumerate}
\end{remark}

{\samepage
\begin{theorem}
\label{thm:inclusions}
\begin{enumerate}
\item \label{thm:inclusions-1} $\mbox{FINITE} \seq \easyforall \seq
  \easyforallio \seq \easyexistsio \seq \np$.

\item \label{thm:inclusions-2} $\easyforall \seq \easyexists \seq
  \easyexistsio$.

\item \label{thm:existsae} $\easyexists = \p$.

\item \label{thm:existsio} $\easyexistsio = 
  \overline{\mbox{P-immune}} \cap \np$.
\end{enumerate}
\end{theorem}
} %

\begin{figure}
\begin{center}
\parbox{10.4cm}{
\setlength{\unitlength}{0.00083300in}%
\begingroup\makeatletter\ifx\SetFigFont\undefined
\def\x#1#2#3#4#5#6#7\relax{\def\x{#1#2#3#4#5#6}}%
\expandafter\x\fmtname xxxxxx\relax \def\y{splain}%
\ifx\x\y   
\gdef\SetFigFont#1#2#3{%
  \ifnum #1<17\tiny\else \ifnum #1<20\small\else
  \ifnum #1<24\normalsize\else \ifnum #1<29\large\else
  \ifnum #1<34\Large\else \ifnum #1<41\LARGE\else
     \huge\fi\fi\fi\fi\fi\fi
  \csname #3\endcsname}%
\else
\gdef\SetFigFont#1#2#3{\begingroup
  \count@#1\relax \ifnum 25<\count@\count@25\fi
  \def\x{\endgroup\@setsize\SetFigFont{#2pt}}%
  \expandafter\x
    \csname \romannumeral\the\count@ pt\expandafter\endcsname
    \csname @\romannumeral\the\count@ pt\endcsname
  \csname #3\endcsname}%
\fi
\fi\endgroup
\begin{picture}(4844,3944)(3579,-6383)
\thicklines
\put(3601,-5761){\line( 1, 0){4800}}
\put(3601,-5161){\line( 1, 0){4800}}
\put(3601,-2461){\line( 1, 0){4800}}
\put(3601,-3361){\line( 1, 0){4800}}
\put(3601,-2461){\line( 0,-1){3900}}
\put(3601,-6361){\line( 1, 0){4800}}
\put(8401,-6361){\line( 0, 1){3900}}
\put(3601,-3661){\line( 4,-1){4800}}
\put(3601,-4861){\line( 4, 1){4800}}
\put(4801,-3661){\makebox(0,0)[lb]{\smash{\SetFigFont{12}{14.4}{rm}$\easyexistsio = \overline{\mbox{P-immune}} \cap \np$}}}
\put(5626,-6061){\makebox(0,0)[lb]{\smash{\SetFigFont{12}{14.4}{rm}FINITE}}}
\put(5626,-5461){\makebox(0,0)[lb]{\smash{\SetFigFont{12}{14.4}{rm}$\easyforall$}}}
\put(5851,-2761){\makebox(0,0)[lb]{\smash{\SetFigFont{12}{14.4}{rm}NP}}}
\put(3751,-4336){\makebox(0,0)[lb]{\smash{\SetFigFont{12}{14.4}{rm}$\easyexists = \p$}}}
\put(7651,-4336){\makebox(0,0)[lb]{\smash{\SetFigFont{12}{14.4}{rm}$\easyforallio$}}}
\end{picture}
} %
\end{center}
\caption{\label{fig:inclusions} Inclusions between classes of NP sets 
having easy certificates.}
\end{figure}
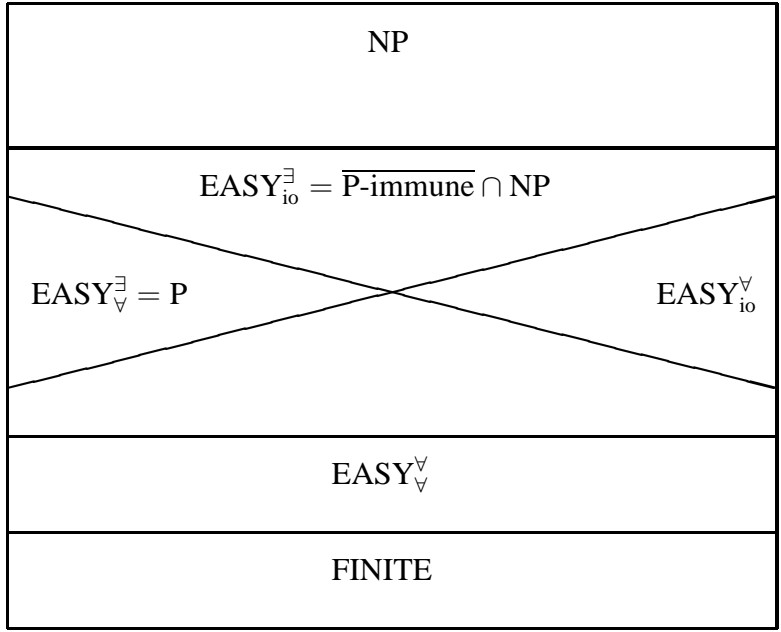

\noindent
{\bf Proof.}
\quad
(\ref{thm:inclusions-1}) \& (\ref{thm:inclusions-2}) The inclusions
immediately follow from Definition~\ref{def:easy}.

\smallskip

(\ref{thm:existsae}) The inclusion $\p \seq \easyexists$ holds by
definition. The converse inclusion $\easyexists \seq \p$ also is
clear, since if $L$ is any set in $\easyexists$ and this is witnessed
by NPM $N$ and FP function $f_{N}$, then there exists a DPM $M$ that
recognizes $L$ as follows. On input $x$, $M$ simulates that
computation path of $N(x)$ that is printed by $f_{N}(x)$.  If $x \in
L$, then $f_{N}(x) \in \mbox{acc}_{N}(x)$, and $M$ accepts $x$. If $x
\not\in L$, then $f_{N}(x)$ cannot be an accepting path of $N(x)$, and
thus $M$ rejects $x$.

\smallskip

(\ref{thm:existsio}) Let $L \in \easyexistsio$ via NPM $N$ and FP
function $f_{N}$. Clearly, $L \in \np$, and if $L$ is not a finite
set, then the set
\[
\{ x \in L \,|\,  \mbox{$f_{N}(x) = y$ and $N(x)$ accepts on path $y$ } \}
\]
is an infinite subset of $L$ that is in P\@. Hence, $L$ is not
P-immune.

Conversely, let $A$ be any NP set that is not P-immune. Let $M_A$ be
an NPM accepting $A$. If $A$ is finite, then we are done. So suppose
$A$ is infinite, and there is an infinite set $B$ such that $B \seq A$
and $B \in \p$ via DPM $M_B$. We now describe an NPM $N$ and an FP
function $f_{N}$ that witness $A \in \easyexistsio$. On input $x$, $N$
first simulates the computation of $M_{B}(x)$, and accepts $x$ if
$M_B$ accepts $x$. If $M_B$ rejects $x$, then $N$ simulates the
computation of $M_{A}(x)$. Clearly, $L(N) = A$. $f_{N}(x)$ is defined
to be the (suitably encoded) computation of $M_{B}(x)$. Since $B$ is
an infinite set, $f_{N}(x)$ prints an accepting path of $N(x)$ for
infinitely many $x \in A$.  
\hfill$\Box$

\medskip

The inclusion relations between FINITE, NP, and all four classes of
easy NP sets ($\easyforall$, $\easyexists$, $\easyforallio$, and
$\easyexistsio$)
are displayed in Figure~\ref{fig:inclusions}.

The Kolmogorov complexity of finite strings was introduced
independently by Kolmogorov~\cite{kol:j:information} and
Chaitin~\cite{cha:j:length-of-programs}. Roughly speaking, the
Kolmogorov complexity of a finite binary string $x$ is the length of a
shortest program that generates $x$. Intuitively, if a string $x$ can
be generated by a program shorter than $x$ itself, then $x$ can be
``compressed.'' The notion of {\em generalized\/} 
Kolmogorov
complexity~(\cite{adl:t:time-randomness,har:c:generalized-kolmogorov,sip:c:randomness}, see the paper of Li and Vit\'{a}nyi~\cite{li-vit:bselman:kol} 
for a nice survey of the field)
is a version of Kolmogorov complexity that provides information about
not only whether and how far a string can be compressed, but also how
fast it can be ``restored.'' We now give the definition of
(unconditional and conditional) generalized Kolmogorov complexity.

\begin{definition}
\label{def:gen-kolmogorov-compl}
(\cite{har:c:generalized-kolmogorov}, see
also~\cite{adl:t:time-randomness,sip:c:randomness})
\quad
For any Turing machine $T$ and functions~$s$ and~$t$ mapping~$\N$ 
to~$\N$, define
\[
\mbox{K}_T[s(n),t(n)] \equalsdef \{ x
\,|\, (\exists y)\, [ \, |x| = n \ \mbox{and}\ |y| \leq s(n) \ 
\mbox{and $T(y)$ outputs $x$ in at most $t(n)$ steps}\, ]\}.
\]
\end{definition}

It was shown in~\cite{har:c:generalized-kolmogorov} that there exists
a {\em universal\/}\footnote{%
\protect\singlespacing
  Roughly speaking, a universal Turing machine $U$ expects as input a
  pair of a (suitably encoded) Turing machine $T$ and an input string
  $y$ and simulates the computation of $T(y)$.  More precisely,
  denoting the encoding of $T$ by $\mbox{code}(T)$ and using our pairing
  function, $U$ runs on input $\< \mbox{code}(T), y \>$ and outputs the
  result of the computation of $T(y)$.
}
Turing machine $U$ such that for any other Turing machine $T$ there
exists a constant $c$ such that
\[
\mbox{K}_T[s(n),t(n)] \seq \mbox{K}_U[s(n) + c,\, c \, t(n) \log t(n) +
c].
\] 
Fixing a universal Turing machine $U$ and dropping the subscript, the
{\em unconditional\/} generalized Kolmogorov complexity will be
denoted by \mbox{$\mbox{K}[s(n),t(n)]
  \equalsdef \mbox{K}_U[s(n),t(n)]$}.
The {\em conditional\/} generalized Kolmogorov complexity (under
condition $z$), in which the information of the string $z$ is given
for free and does not count for the complexity, is defined as follows.

\begin{definition}
\label{def:cond-kolmogorov-compl}
\quad 
Let $U$ be a fixed universal Turing machine and $z$ be a string.  For
functions~$s$ and~$t$ mapping~$\N$ to~$\N$, define
\[
\mbox{K}[s(n),t(n) \condition z] \equalsdef \{ x \,|\, (\exists y)\, [
\, |x| = n \ \mbox{and} \ |y| \leq s(n) \ \mbox{and $U(\< y, z \>)$
  outputs $x$ in $\leq t(n)$ steps}\, ] \}.
\]
\end{definition}

In particular, \mbox{$\mbox{K}[s(n),t(n) \condition \epsilon] =
  \mbox{K}[s(n),t(n)] $}.

Of particular interest in this paper are certificates (more precisely,
strings encoding accepting paths of NPMs) that have {\em small\/}
generalized Kolmogorov complexity.
Levin~(see~\cite{tra:j:per}) and
\mbox{Adleman~\cite{adl:t:time-randomness}} independently discovered
the connection between small generalized Kolmogorov complexity and
certificates.  This connection has also been used in other contexts
(\cite{hem-wec:j:man-rand},
see also
\cite{hem-rud:j:ranking,gre-tor:unpub:kol} and the comments
in~\cite{hem-rud:j:ranking} on~\cite{cai-hem:j:enum}).

\begin{definition}
\label{def:p-printable}
\cite{har-yes:j:computation-times}
\quad
A set $S$ is P{\em -printable\/} if there exists a DPM $M$ such that
for each length~$n$, $M$ on input~$1^n$ prints all elements of~$S$
having length at most~$n$.
\end{definition}

The P-printable sets are closely related to sets of strings having
small unconditional generalized Kolmogorov complexity: A set $S$ is
P-printable if and only if $S \in \p$ and $S \seq \mbox{K}[k \log n,
n^k]$ for some constant~$k$ (\cite{all-rub:j:print}, see
also~\cite{bal-boo:j:kolmogorov,har-hem:j:sparse-oracles,rub:t:small-kolmogorov}).
Below we note a similar connection between the sets in $\easyforall$
and $\easyforallio$ and the sets of certificates having small {\em
  conditional\/} generalized Kolmogorov complexity, thus showing the
robustness of these notions.  Due to Theorem~\ref{thm:inclusions}, the
corresponding claims for $\easyexists$ and $\easyexistsio$ are omitted.
Though the flavor 
of the correspondence here invoked is by now standard (e.g., see the 
above papers), 
we include the proof 
of Observation~\ref{NOWobs:WW} for completeness.

\begin{observation} \label{NOWobs:WW} 
\quad The following are equivalent:
\begin{enumerate}
\item $L \in \easyforall$.

\item For each normalized NPM $N$ accepting $L$ there is 
  a constant $k$ (which may depend on $N$) such that for each string
  $x \in L$ it holds that 
  \mbox{$\mbox{acc}_{N}(x) \cap \mbox{K}[ k \log n, n^k \condition x] \neq
  \emptyset$}.
\end{enumerate}
\end{observation}

\noindent
{\bf Proof.}
\quad
In one direction the function proving a machine easy itself yields
Kolmogorov-simple certificates. That is, for any normalized NPM $N$
accepting the given $\easyforall$ set~$L$, there is an FP function
$f_N$ that outputs an accepting path of $N(x)$ for each~$x \in L$.
Thus, for each~$x \in L$, the certificate $f_N(x)$ is in $\mbox{K}[ k
\log n, n^k \condition x]$ for some constant $k$ depending only on
$f_N$ (and thus on~$N$), since the program for~$f_N$, encoded as a
string~$y$, has constant size, and the universal Turing machine $U$
running on input $\<y, x\>$ can clearly generate $f_N(x)$ in time
polynomial in $|f_N(x)|$.

\smallskip

In the other direction, let $N$ be any NPM accepting~$L$. By
assumption, for each~$x \in L$, $N(x)$ has certificates of small
conditional Kolmogorov complexity relative to $x$ (i.e., it has
certificates in $\mbox{K}[ k \log n, n^k \condition x]$ for some
constant~$k$).  
Note that $n$, the length of those certificates, is
polynomial in~$|x|$;  let $p$ be some such 
polynomial bound.
So, for each~$x$, $n = p(|x|)$ is a polynomial bound
on the length of the certificates of~$N(x)$.  There are at most
$2^{{\cal O} (\log n)} = n^{\ell}$ (for some suitable constant~$\ell$)
short strings that potentially encode programs $y$ such that the
universal Turing machine~$U$, running on input $\<y, x\>$, produces a
certificate of $N(x)$ in time polynomial in $n$, say in time~$n^m =
(p(|x|))^{m}$. Let \mbox{$q(|x|) \equalsdef \max \{(p(|x|))^{\ell},
  (p(|x|))^{m} \}$}.

The function $f_N$ proving $N$ easy works on input $x$ as follows. In
a brute-force manner, $f_N$ runs the universal machine on the pairs
$\<y,x\>$ for all the at most $q(|x|)$ many short strings~$y$, $|y|
\leq k \log n$, for $q(|x|)$ steps, and then for each output
checks if the output is an accepting path of~$N(x)$, and eventually
outputs the first such accepting path found.  If no accepting path was
found, the input $x$ is not in~$L$. Clearly, $f_N$ is polynomial-time
computable and, for each input~$x\in L$, outputs a certificate of
$N(x)$.~\hfill$\Box$

\begin{remark} \quad
  Note that the normalization requirement in the above observation is
  crucial, since our definition of conditional generalized Kolmogorov
  complexity displays the strange feature that for machines that are
  not normalized, if we use a certain simple polynomial-time
  computable and polynomial-time invertible pairing function, say $\<
  \cdot, \cdot \>_{\mbox{\protect\scriptsize weird}}$, to encode the
  pair of the program $y$ and the condition $z$ as input $\< y, z
  \>_{\mbox{\protect\scriptsize weird}}$ to the universal Turing
  machine, then even the empty string has {\em non-constant\/}
  conditional Kolmogorov complexity. Due to our normalization
  requirement, however, this issue is not germane here.
\end{remark}

The proof of Observation~\ref{NOWobs:YY} follows precisely the lines
of the proof of Observation~\ref{NOWobs:WW}.

\begin{observation} \label{NOWobs:YY}  
\quad The following are equivalent:
\begin{enumerate}
\item $L \in \easyforallio$.

\item For each normalized NPM $N$ accepting $L$ there is a constant $k$ 
  (which may depend on $N$) such that for infinitely
  many strings $x \in L$ it holds that 
  \mbox{$\mbox{acc}_{N}(x) \cap \mbox{K}[ k \log n, n^k \condition x] \neq
  \emptyset$}.
\end{enumerate}
\end{observation}

\section{Positive Results}
\label{sec:positive}

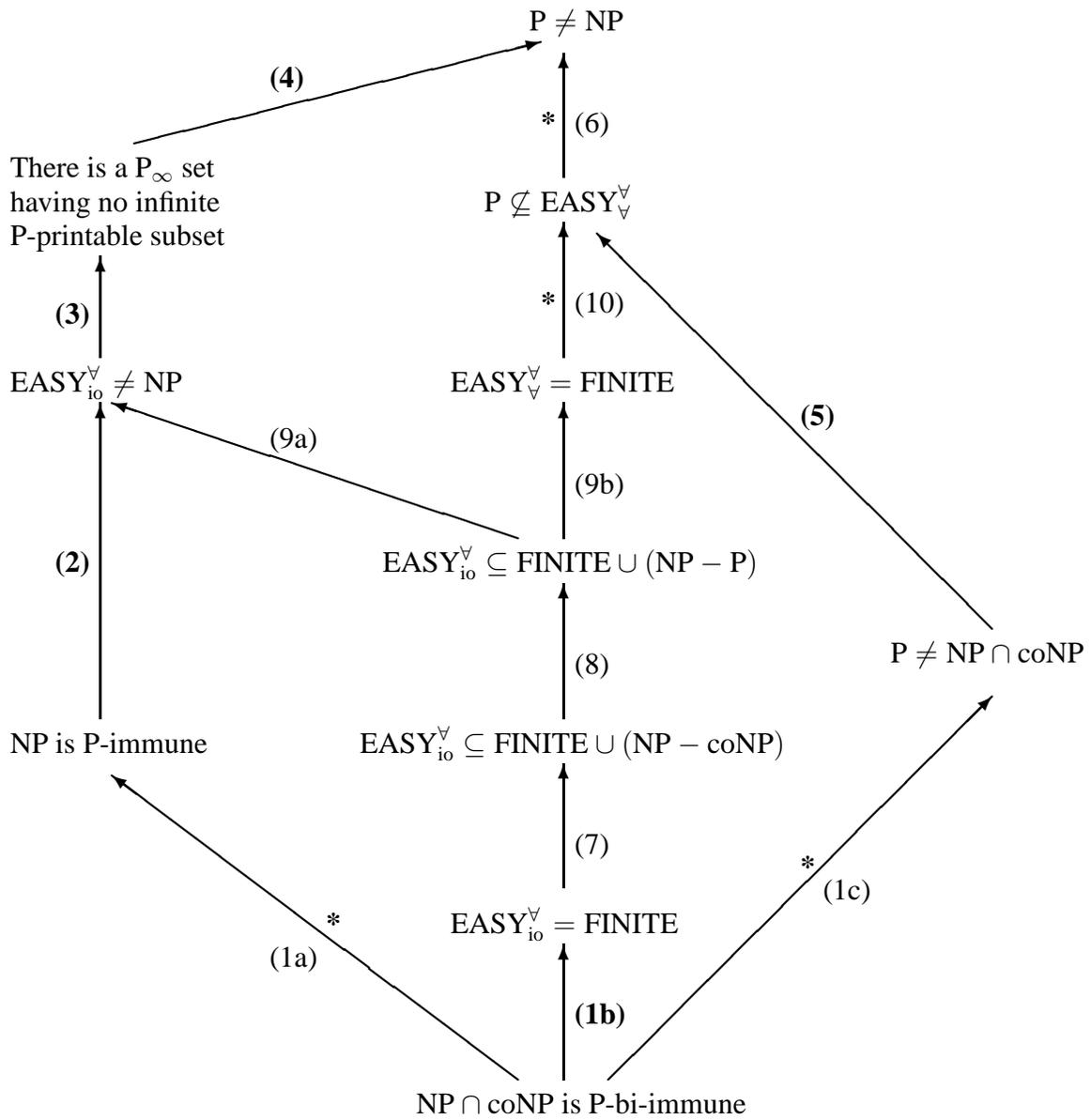
\begin{figure}
\setlength{\unitlength}{0.00083300in}%
\begingroup\makeatletter\ifx\SetFigFont\undefined
\def\x#1#2#3#4#5#6#7\relax{\def\x{#1#2#3#4#5#6}}%
\expandafter\x\fmtname xxxxxx\relax \def\y{splain}%
\ifx\x\y   
\gdef\SetFigFont#1#2#3{%
  \ifnum #1<17\tiny\else \ifnum #1<20\small\else
  \ifnum #1<24\normalsize\else \ifnum #1<29\large\else
  \ifnum #1<34\Large\else \ifnum #1<41\LARGE\else
     \huge\fi\fi\fi\fi\fi\fi
  \csname #3\endcsname}%
\else
\gdef\SetFigFont#1#2#3{\begingroup
  \count@#1\relax \ifnum 25<\count@\count@25\fi
  \def\x{\endgroup\@setsize\SetFigFont{#2pt}}%
  \expandafter\x
    \csname \romannumeral\the\count@ pt\expandafter\endcsname
    \csname @\romannumeral\the\count@ pt\endcsname
  \csname #3\endcsname}%
\fi
\fi\endgroup
\begin{picture}(6547,7386)(2326,-8191)
\thicklines
\put(6001,-3136){\vector( 0, 1){900}}
\put(6001,-4336){\vector( 0, 1){900}}
\put(6001,-5536){\vector( 0, 1){900}}
\put(6001,-6661){\vector( 0, 1){825}}
\put(5701,-4336){\vector(-3, 1){2700}}
\put(5701,-7936){\vector(-4, 3){2700}}
\put(6301,-7936){\vector( 1, 1){2550}}
\put(6001,-1936){\vector( 0, 1){825}}
\put(6001,-7936){\vector( 0, 1){900}}
\put(2926,-5536){\vector( 0, 1){2100}}
\put(2926,-3136){\vector( 0, 1){675}}
\put(3151,-1711){\vector( 4, 1){2700}}
\put(8851,-4936){\vector(-1, 1){2625}}
\put(5026,-8161){\makebox(0,0)[lb]{\smash{\SetFigFont{12}{14.4}{rm}$\np \cap \conp$ is P-bi-immune}}}
\put(2326,-3361){\makebox(0,0)[lb]{\smash{\SetFigFont{12}{14.4}{rm}$\easyforallio \neq \np$}}}
\put(2326,-2161){\makebox(0,0)[lb]{\smash{\SetFigFont{12}{14.4}{rm}having no infinite}}}
\put(2326,-2386){\makebox(0,0)[lb]{\smash{\SetFigFont{12}{14.4}{rm}P-printable subset}}}
\put(5476,-2161){\makebox(0,0)[lb]{\smash{\SetFigFont{12}{14.4}{rm}$\p \not\seq \easyforall$}}}
\put(5776,-961){\makebox(0,0)[lb]{\smash{\SetFigFont{12}{14.4}{rm}$\p \neq \np$}}}
\put(4051,-7186){\makebox(0,0)[lb]{\smash{\SetFigFont{12}{14.4}{rm}(1a)}}}
\put(7726,-6736){\makebox(0,0)[lb]{\smash{\SetFigFont{12}{14.4}{rm}(1c)}}}
\put(6076,-2836){\makebox(0,0)[lb]{\smash{\SetFigFont{12}{14.4}{rm}(10)}}}
\put(6076,-4036){\makebox(0,0)[lb]{\smash{\SetFigFont{12}{14.4}{rm}(9b)}}}
\put(6076,-5236){\makebox(0,0)[lb]{\smash{\SetFigFont{12}{14.4}{rm}(8)}}}
\put(6076,-6436){\makebox(0,0)[lb]{\smash{\SetFigFont{12}{14.4}{rm}(7)}}}
\put(2626,-4561){\makebox(0,0)[lb]{\smash{\SetFigFont{12}{14.4}{bf}(2)}}}
\put(2626,-2911){\makebox(0,0)[lb]{\smash{\SetFigFont{12}{14.4}{bf}(3)}}}
\put(4051,-1336){\makebox(0,0)[lb]{\smash{\SetFigFont{12}{14.4}{bf}(4)}}}
\put(7576,-3586){\makebox(0,0)[lb]{\smash{\SetFigFont{12}{14.4}{bf}(5)}}}
\put(5251,-6961){\makebox(0,0)[lb]{\smash{\SetFigFont{12}{14.4}{rm}$\easyforallio = \mbox{FINITE}$}}}
\put(4801,-4561){\makebox(0,0)[lb]{\smash{\SetFigFont{12}{14.4}{rm}$\easyforallio \seq \mbox{FINITE} \cup (\np - \p)$}}}
\put(4651,-5761){\makebox(0,0)[lb]{\smash{\SetFigFont{12}{14.4}{rm}$\easyforallio \seq \mbox{FINITE} \cup (\np - \conp)$}}}
\put(6076,-7561){\makebox(0,0)[lb]{\smash{\SetFigFont{12}{14.4}{bf}(1b)}}}
\put(4426,-6961){\makebox(0,0)[lb]{\smash{\SetFigFont{12}{14.4}{bf}*}}}
\put(7576,-6586){\makebox(0,0)[lb]{\smash{\SetFigFont{12}{14.4}{bf}*}}}
\put(8176,-5161){\makebox(0,0)[lb]{\smash{\SetFigFont{12}{14.4}{rm}$\p \neq \np \cap \conp$}}}
\put(5251,-3361){\makebox(0,0)[lb]{\smash{\SetFigFont{12}{14.4}{rm}$\easyforall = \mbox{FINITE}$}}}
\put(6076,-1636){\makebox(0,0)[lb]{\smash{\SetFigFont{12}{14.4}{rm}(6)}}}
\put(5851,-1636){\makebox(0,0)[lb]{\smash{\SetFigFont{12}{14.4}{bf}*}}}
\put(2326,-5761){\makebox(0,0)[lb]{\smash{\SetFigFont{12}{14.4}{rm}NP is P-immune}}}
\put(2326,-1936){\makebox(0,0)[lb]{\smash{\SetFigFont{12}{14.4}{rm}There is a $\pinf$ set}}}
\put(4051,-3736){\makebox(0,0)[lb]{\smash{\SetFigFont{12}{14.4}{rm}(9a)}}}
\put(5851,-2836){\makebox(0,0)[lb]{\smash{\SetFigFont{12}{14.4}{bf}*}}}
\end{picture}
\caption{\label{fig:arrows} Some implications between 
various properties of (classes of) sets within NP.}  
\end{figure}

In this section, we prove a number of implications between certain
properties of subclasses of NP that are summarized in 
Figure~\ref{fig:arrows}.
Usually, when one is trying to give strong evidence for some 
complexity-theoretic statement
$A$ not to be true, one does so by showing that $A$ implies~$\p =
\np$. In contrast, our Figure~\ref{fig:arrows} has $\p \neq \np$ as
its top conclusion. Nonetheless, the implications of
Figure~\ref{fig:arrows} are not meaningless. Their importance is obvious 
in light of the fact that
the statements of the figure (in particular, the immunity assertions
and $\p \neq \np \cap \conp$ as well as the condition $\p \neq
\easyforall$, which is equivalent to the existence of surjective
one-way functions (see~\cite{fen-for-nai-rog:c:inverse})) are
well-studied and important conditions in complexity theory. The
implications of Figure~\ref{fig:arrows} simply state that these
conditions are at least as hard to prove as proving~$\p \neq \np$, and
they explore the logical fine-structure amongst those important conditions.

Here is the key for Figure~\ref{fig:arrows}: Implications
represented by arrows that are marked by a ``*'' are not invertible up
to the limits of relativizations (as will be shown in
Section~\ref{sec:negative}). Consequently, no chain of implications
that contains an arrow marked by a ``*'' is invertible in all
relativized worlds.  Arrows labeled by boldface numbers indicate
non-trivial implications to be proven in Theorem~\ref{thm:arrows}.

We first discuss the trivial implications of Figure~\ref{fig:arrows}. 
We stress that these trivial
statements are included not only in order to make the picture
displayed in Figure~\ref{fig:arrows} as complete as possible, but also
for the following reason. In the next section, we will prove that the
reverse of some implication chains comprising both trivial and
more interesting implications (the latter ones being stated in
Theorem~\ref{thm:arrows} below) not only fails in some relativized
worlds, but, even worse, this relativized failure can already be shown
for the trivial part of the implication chain considered.  Therefore, 
it does make sense to explicitly state such trivial implications.

These trivial facts are either immediately clear, or they follow from
simple set-theoretical arguments, or are straightforwardly established
by the equivalences given in Proposition~\ref{thm:equivalences} below. For
instance, the
equivalence of 
``\mbox{$\easyforall = \mbox{FINITE}$}'' and ``\mbox{$\easyforall \seq
  \mbox{FINITE} \cup (\np - \p)$}'' can be seen by simple
set-theoretical considerations.\footnote{%
\protect\singlespacing
  To be definite, for all sets $A$, $B$, $C$, and~$X$, if $A \seq X \seq B
  \seq C$, then $(X = A \Lolra X \seq A \cup
  (C - B))$\@.  Taking \mbox{$A = \mbox{FINITE}$}, \mbox{$B = \p$},
  \mbox{$C = \np$}, and~\mbox{$X = \easyforall$}, we have verified our claim.
}
The statement ``\mbox{$\easyforall = \mbox{FINITE}$},'' in turn, 
immediately implies 
the statement ``\mbox{$\p \not\seq \easyforall$}'' (see arrow (10) in
Figure~\ref{fig:arrows}). We have been informed that the authors 
of~\cite{fen-for-nai-rog:c:inverse}
have shown a number of very interesting conditions, including
``\mbox{$\sigmastar \not\in \easyforall$}'' and ``there exists an honest
polynomial-time computable {\em onto\/} function that is not 
polynomial-time invertible,'' 
to be all equivalent to
the statement ``\mbox{$\p \not\seq \easyforall$}.''

Furthermore, the arrows in Figure~\ref{fig:arrows} labeled
(1a), (1c), (7), and~(8) are immediately clear. Concerning the
arrows
(9a) and~(9b), note that (9b) follows from the equivalence of
``\mbox{$\easyforall = \mbox{FINITE}$}'' and ``\mbox{$\easyforall \seq
  \mbox{FINITE} \cup (\np - \p)$}'' stated in the previous paragraph,
whereas (9a) is implied by
Proposition~\ref{thm:equivalences}.\ref{thm:equivalences-3} below.
Similarly, 
arrow (6) holds due to
Proposition~\ref{thm:equivalences}.\ref{thm:equivalences-1}, since if
\mbox{$\p \not\seq \easyforall$}, then there exists a set in
\mbox{$\np - \easyforall$}, and thus we have \mbox{$\p \neq \np$}\@.

The following proposition gives characterizations 
for two nodes of Figure~\ref{fig:arrows}.

\begin{proposition}
\label{thm:equivalences}
\begin{enumerate}
\item \label{thm:equivalences-1} $\p \neq \np$ \ if and only if \ 
  $\easyforall \neq \np$.

\item \label{thm:equivalences-3} $\easyforallio \seq \mbox{FINITE}
  \cup (\np - \p)$ \ if and only if \ $\sigmastar \not\in
  \easyforallio$.
\end{enumerate}
\end{proposition}

\noindent
{\bf Proof.}
\quad
(\ref{thm:equivalences-1})
Adleman (\cite{adl:t:time-randomness}, see 
also~\cite{tra:j:per} for a discussion of 
Levin's related work)
has
shown that $\p = \np$ if and only if for each normalized NPM $M$ 
there is a $k$ such that 
for each string $x \in L(M)$ it holds that 
\mbox{$\mbox{acc}_{M}(x) \cap \mbox{K}[ k
\log n, n^k \condition x] \neq \emptyset$}.
By
Observation~\ref{NOWobs:WW}, this implies that $\p = \np$ if and only if
$\easyforall = \np$.

\smallskip

(\ref{thm:equivalences-3}) First note that the statement
``\mbox{$\easyforallio \seq \mbox{FINITE} \cup (\np - \p)$''} is
equivalent to ``\mbox{$\easyforallio \cap \pinf = \emptyset$},'' and
thus immediately implies $\sigmastar \not\in \easyforallio$, since
clearly $\sigmastar \in \pinf$. For the converse implication, assume
there exists a set $L$ in $\easyforallio \cap \pinf$. Let $M_L$ be a
DPM such that $L(M_{L}) = L$. We show that $\sigmastar \in
\easyforallio$. Let $N$ be any NPM such that $L(N) = \sigmastar$.  By
way of contradiction, suppose $N$ has easy certificates only for
finitely many $x \in \sigmastar$. Consider the following NPM $N_L$ for
$L$. On input $x$, $N_L$ first simulates the computation of $N(x)$,
and then, for every path of this simulation, $N_L$ simulates $M_L(x)$
and accepts accordingly. Clearly, $L(N_L) = L$. However, by our
supposition that $N$ has easy certificates for finitely many inputs
only, $N_L$ also can have easy certificates for at most finitely many
inputs, contradicting that $L \in \easyforallio$.  Thus, $\sigmastar
\in \easyforallio$. This completes the proof.~\hfill$\Box$

\medskip

Next, we prove the non-trivial implications of Figure~\ref{fig:arrows}.

\begin{theorem}
\label{thm:arrows}
\begin{enumerate}
\item \label{thm:arrows-1} If $\np \cap \conp$ is P-bi-immune, then
$\easyforallio = \mbox{FINITE}$.

\item \label{thm:arrows-2} If NP is P-immune, then $\easyforallio \neq
  \np$.

\item \label{thm:arrows-3} If $\easyforallio \neq \np$, then there
  exists an infinite P set having no infinite P-printable subset.

\item \label{thm:arrows-4} \cite{all:b:kol} \quad 
  If there exists an infinite P set having no
  infinite P-printable subset, then $\p \neq \np$.

\item \label{thm:arrows-5} \cite{bor-dem:t:selfreducibility} \quad If
  $\np \cap \conp \neq \p$, then $\p \not\seq \easyforall$.

\end{enumerate}
\end{theorem}

\noindent
{\bf Proof.} 
\quad
(\ref{thm:arrows-1}) Let $Q$ be any P-bi-immune set such that $Q \in
\np \cap \conp$ via NPMs $N_{Q}$ and $N_{\overline{Q}}$, that is,
$L(N_{Q}) = Q$ and $L(N_{\overline{Q}}) = \overline{Q}$.  By way of
contradiction, assume there exists an infinite set $L$ in $\easyforallio$. 
Let $N$ be any NPM accepting~$L$.
Consider the following NPM $\widehat{N}$ for~$L$.  Given~$x$,
$\widehat{N}$ runs $N(x)$ and rejects on all rejecting paths
of~$N(x)$.  On all accepting paths of~$N(x)$, $\widehat{N}$
nondeterministically guesses whether $x \in Q$ or $x \in
\overline{Q}$, simultaneously guessing certificates (i.e., accepting
paths of $N_{Q}(x)$ or~$N_{\overline{Q}}(x)$) for whichever guess was
made, and accepts on each accepting path of $N_{Q}(x)$
or~$N_{\overline{Q}}(x)$. Clearly, $L(\widehat{N}) = L$\@. By our
assumption that $L$ is an infinite set in $\easyforallio$,
$\widehat{N}$ has easy certificates for infinitely many inputs. Let
$f_{\widehat{N}}$ be an FP function that infinitely often outputs an
easy certificate of $\widehat{N}$.  Let
\[
\widehat{L} \equalsdef \{ x \,|\,
f_{\widehat{N}}(x) \ \mbox{outputs an easy certificate of} \
\widehat{N}(x)\}.
\]
Note that $\widehat{L}$ is an infinite subset of $L$, and that for any
input~$x$, it can be checked in polynomial time whether $x$ belongs to
$Q \cap \widehat{L}$ or $\overline{Q} \cap \widehat{L}$, respectively,
by simply checking whether the string printed by $f_{\widehat{N}}$
indeed certifies either $x \in Q \cap \widehat{L}$ or $x \in
\overline{Q} \cap \widehat{L}$. Thus, either $Q \cap \widehat{L}$ or
$\overline{Q} \cap \widehat{L}$ must be an infinite set in P, which
contradicts that $Q$ is P-bi-immune. Hence, every set in
$\easyforallio$ is finite.

\smallskip

(\ref{thm:arrows-2}) Let $L$ be any P-immune NP set. We claim that $L
\not\in \easyforallio$. Suppose to the contrary that $L \in
\easyforallio$. Let $N$ be any NPM accepting $L$. Then there exists an
FP function $f_N$ such that $f_{N}(x) \in \mbox{acc}_{N}(x)$ for
infinitely many inputs $x$. Define
\[
B \equalsdef \{ x \,|\, f_{N}(x) \in
\mbox{acc}_{N}(x)\}.
\]
Then, $B$ is an infinite subset of $L$ and $B \in \p$, contradicting
the P-immunity of $L$.

\smallskip

(\ref{thm:arrows-3}) If $\easyforallio \neq \np$, then there exist an
infinite NP set $L$ and an NPM $N$ accepting $L$ such that, 
\[
(*)\hspace*{1cm} (\forall f \in \fp)\, (\forallae x \in L )\,
[f(x) \not\in \mbox{acc}_{N}(x)].
\]
Let $q$ be a polynomial such that $|\< x, y \>| = q(|x|)$ for any
string $x$ and any path $y$ of $N(x)$. Define
\[
D \equalsdef \{ \< x, y \> \,|\, y \in
\mbox{acc}_{N}(x)\}.
\]
Clearly, $D$ is an infinite set in P\@. Suppose there exists an
infinite set $A$ such that $A \seq D$ and $A$ is P-printable via some
DPM $M$\@. Define an FP function $f_A$ that is computed by DPM $M_A$
as follows.  On input $x$, $M_A$ simulates the computation of
$M(1^{q(|x|)})$ and prints all elements of $A$ up to length~$q(|x|)$.
If a string of the form $\< x, y \>$ is printed, $M_A$ outputs $y$.
Clearly, $f_{A}(x) \in \mbox{acc}_{N}(x)$ for infinitely many $x \in
L$, contradicting $(*)$ above. Hence, $D$ has no infinite P-printable
subset.

\smallskip

(\ref{thm:arrows-4}) This implication can be seen from results due to
Allender~\cite{all:b:kol}. First, some definitions are needed. Let us
consider another version of time-bounded Kolmogorov complexity, a
version that is due to Levin~\cite{lev:j:randomness} (see
also~\cite{lev:j:universal}).  For the fixed universal Turing machine
$U$ and any string~$x$, define $\mbox{Kt}(x)$ to be
\[
\min\{|y|+\log n \,|\, U(y) \mbox{ outputs $x$ in at most $n$ steps } \}.
\]
For any set~$L$, let $\mbox{K}_{L}(n) \equalsdef \min\{\mbox{Kt}(x)
\,|\, x \in L^{=n} \}$. As is standard, let E and NE denote
respectively $\bigcup_{c > 0\:} {\rm DTIME}(2^{cn})$ and $\bigcup_{c >
  0\:} {\rm NTIME}(2^{cn})$. An {\em NE predicate\/} is a relation $R$
defined by an NE machine~$M$: $R(x,y)$ is true if and only if $y$
encodes an accepting path of $M(x)$. An NE predicate $R$ is {\em
  E-solvable\/} if there is some function $f$ computable in time
$2^{cn}$ for some constant~$c$ such that \mbox{$(\forall x)\,
  [(\exists y)\, [R(x,y)] \Lolra R(x,f(x))]$}.

In~\cite{all:b:kol}, Allender proves that (a)~there exists an infinite
P set having no infinite P-printable subset if and only if there
exists a set $B \in \p$ such that $\mbox{K}_{B}(n) \in \omega(\log
n)$, and (b)~there exists an NE predicate that is not E-solvable if
and only if there exists a set $C \in \p$ such that $\mbox{K}_{C}(n)
\not\in {\cal O}(\log n)$. Since $\mbox{K}_{B}(n) \in \omega(\log n)$
clearly implies $\mbox{K}_{B}(n) \not\in {\cal O}(\log n)$ and since
the existence of an NE predicate that is not E-solvable implies $\p
\neq \np$, (\ref{thm:arrows-4}) is proven. 

For completeness and to enhance readability, we add a more transparent
direct proof of~(\ref{thm:arrows-4}).  To prove the contrapositive,
assume $\p = \np$.  Let $L$ be any infinite set in P\@. Define the set
\[
A \equalsdef \{ \< 0^n, w\> \,|\, n \geq 0 \,\wedge\, |w| = n
\,\wedge\, (\exists z \in L^{=n})\, 
[ z <_{\mbox{\protect\scriptsize lex}} w] \}.
\]
Clearly, $A \in \np$, and by our assumption, $A \in \p$. Define the
set of the lexicographically smallest length~$n$ strings of~$L$ for
each length~$n$:
\[
S \equalsdef \{ x \in L \,|\,
(\forall y \in L^{=|x|})\, [x \leq_{\mbox{\protect\scriptsize lex}} y]\}.
\]
Clearly, $S$ is an infinite subset of $L$. Furthermore, $S$ is
P-printable, since we can find, at each length~$n$, the
lexicographically smallest length~$n$ string in $L$ (which is the
length~$n$ string of $S$) via prefix search that can be performed in
$\fp^{A} = \fp$\@. Thus, every infinite set in P has an infinite
P-printable subset, as was to be shown.

\smallskip 

(\ref{thm:arrows-5}) The proof of this result is implicit in the most
common proof (often credited as Hartmanis's simplification of the
proof of Borodin and Demers) of the theorem of Borodin and
Demers~\cite{bor-dem:t:selfreducibility}, here stated as
Theorem~\ref{thm:bor-dem}
(see~\cite{sel:j:natural} for related work bearing upon the theorem 
of Borodin and Demers),
as has been noted independently of the present paper by 
Fenner et al.~\cite{fen-for-nai-rog:c:inverse}.
For completeness, we
include the proof that $\np \cap \conp \neq \p$ implies $\p \not\seq
\easyforall$. Let $L \in \np \cap \conp$ via NPMs $N_{L}$ and
$N_{\overline{L}}$, that is, \mbox{$L(N_{L}) = L$} and
\mbox{$L(N_{\overline{L}}) = \overline{L}$}.  Assume further that $L
\not\in \p$. Consider the following NPM $M$. On input $x$, $M$
nondeterministically guesses whether $x \in L$ or $x \in
\overline{L}$, simultaneously guessing certificates (i.e., accepting
paths of $N_{L}(x)$ or $N_{\overline{L}}(x)$) for whichever guess was
made. Now, \mbox{$L(M) = L(N_{L}) \cup L(N_{\overline{L}}) = L \cup
  \overline{L} = \sigmastar$}.  Clearly, $\sigmastar \in \p$.  We
claim that (under our assumption that $L \not\in \p$) $\sigmastar
\not\in \easyforall$. Suppose to the contrary that \mbox{$\sigmastar
  \in \easyforall$}. Then, for the NPM~$M$ accepting~$\sigmastar$,
there exists an FP function $f_M$ that prints an accepting path of
$M(x)$ on each input~$x$. Hence, $L$ can be decided in polynomial time
by simply checking which path of the initial branching of $M(x)$ led
to acceptance. That is, a DPM for~$L$, on input~$x$, computes
$f_{M}(x)$ and then checks whether the initial nondeterministic guess
of $M(x)$ on the path printed by $f_{M}(x)$ was either $x \in L$ or
$x \in \overline{L}$, and accepts accordingly. 
This contradicts our assumption that $L \not\in \p$. Hence,
$\sigmastar \not\in \easyforall$.  
\hfill$\Box$

\medskip

Finally, we state an interesting observation by Selman. Recall that
$\p = \easyforall$ if and only if $\Sigma^* \in \easyforall$.  The
following claim gives further characterizations of $\p = \easyforall$
in terms of the question of whether $\easyforall$ is closed under
complementation. 

\begin{claim}
\label{sel:perscomm}
\cite{sel:perscomm} 
\quad 
The following are equivalent.
\begin{enumerate}
\item $\p = \easyforall$.

\item $\easyforall$ is closed under complementation.

\item There exists a set $L$ in P such that $L \in \easyforall$ and
  $\overline{L} \in \easyforall$.
\end{enumerate}
\end{claim}

\noindent
{\bf Proof.} 
\quad
Clearly, Statement~(1) implies~(2) and~(2) implies~(3).  Assume $L \in
\p$, $L \in \easyforall$, and $\overline{L} \in \easyforall$. Let $M_1$ 
(respectively, $M_2$) be a DPM that accepts $L$ (respectively,~$\overline{L}$).
Let $N$ be an NPM that accepts~$\Sigma^*$.  Define NPM~$N_1$ so that
on input~$x$, $N_1$ simultaneously simulates $N$ and~$M_1$, and $N_1$
accepts if and only if both $N$ and $M_1$ accept.  Observe that every
accepting computation of $N_1$ encodes an accepting computation of~$N$.  
Similarly, define $N_2$ to simultaneously simulate $N$ and~$M_2$.  
Then, $L(N_1) = L$ and $L(N_2) = \overline{L}$.  Thus, there
exist $f_1$ and $f_2$ in FP such that $x \in L$ implies $f_1(x)$ is an
accepting computation of~$N_1$, and $x \in \overline{L}$ implies
$f_2(x)$ is an accepting computation of~$N_2$.
Define $f(x) = f_1(x)$ if $x \in L$, and $f(x) = f_2(x)$ if $x \in
\overline{L}$.  Then, $f \in \fp$ and for all~$x$, $f(x)$ contains an
encoding of a computation of $N$ on~$x$.  Thus, $\Sigma^* \in
\easyforall$.  
\hfill$\Box$

\medskip

Consider the reverse of arrow~(10) in Figure~\ref{fig:arrows}, i.e.,
the question of whether $\p \not\subseteq \easyforall$ implies
$\easyforall = \mbox{FINITE}$.  Suppose not.  That is, suppose that
$\p \neq \easyforall \neq \mbox{FINITE}$.  Then, there is a set $L$
such that $L$ is infinite, $\overline{L}$ is infinite, $L \in \p$, $L
\in \easyforall$, and $\overline{L} \not\in \easyforall$.  In
Corollary~\ref{cor:nifty} below, we will give an oracle relative
to which $\p \neq \easyforall \neq \mbox{FINITE}$. Since
Claim~\ref{sel:perscomm} and the above comments relativize, in this
world, such a set $L$ indeed exists.

\section{Negative Results}
\label{sec:negative}

In this section, we show that some of the results from the previous
section are optimal with respect to relativizable techniques. That is,
for some of the implications displayed in Figure~\ref{fig:arrows}, we
construct an oracle relative to which the reverse of that implication
does not hold.  For instance, from Parts~(\ref{thm:arrows-2})
and~(\ref{thm:arrows-5}) of Theorem~\ref{thm:arrows} and the trivial
facts that are shown as arrows (1a) and (1c) in
Figure~\ref{fig:arrows}, we have the following implication chains:
\begin{enumerate}
\item If $\np \cap \conp$ is P-bi-immune, then $\np$ is P-immune,
  which in turn implies that \mbox{$\easyforallio \neq \np$}, and

\item If $\np \cap \conp$ is P-bi-immune, then $\np \cap \conp \neq
  \p$, which in turn implies that \mbox{$\p \not\seq \easyforall$}.
\end{enumerate}

First, we prove that the reverse of these chains fails to hold in some
relativized world, and, even worse, that this relativized failure can
be shown via proving that not even the trivial parts of the chains are
invertible for all oracles.  For both chains, this result can be
achieved via one and the same oracle to be constructed in the proof of
Theorem~\ref{thm:cute} below. This relativized world will additionally
satisfy that the inequalities \mbox{$\mbox{FINITE} \neq \easyforall
  \neq \p \neq \np$} simultaneously hold in this world (see
Corollary~\ref{cor:nifty}).

One main technical contribution in the proof of Theorem~\ref{thm:cute}
is that we give a novel application of the classic ``wide spacing''
oracle construction technique: We show that this technique {\em
  instantly\/} yields the non-P-bi-immunity of NP relative to some
oracle. The use of the wide spacing technique dates so far back that it
is hard to know for sure where it was first used.  It certainly played
an important role in the important early paper by
Kurtz~\cite{kur:tCITE-ONLY-AS-MANUSCRIPT:failure} (see also the very
early use of wide gaps to facilitate the brute-force computation of
smaller strings employed by Ladner~\cite{lad:j:np-incomplete}, and
also in Baker, Gill, and Solovay's seminal
work~\cite{bak-gil-sol:j:p-np-oracle} and in Rackoff's oracle
constructions for probabilistic and unambiguous polynomial time
classes~\cite{rac:j:rel}).

\begin{theorem}
\label{thm:cute}
\quad 
There exists a recursive oracle $A$ such that
\begin{description}
\item[(a)] $\np^A = \mbox{PSPACE}^A$,

\item[(b)] $\np^A$ is $\p^A$-immune, and

\item[(c)] $\np^A$ is not $\p^A$-bi-immune.
\end{description}
\end{theorem}

\noindent
{\bf Proof.} 
\quad
The oracle $A$ will be $\mbox{QBF} \oplus B$, where QBF is any fixed
PSPACE-complete problem and the set $B$ is constructed in stages, $B
\equalsdef \bigcup_{j \geq 0} B_j$.
Define the function $t$ and the sets $T$ and $T_k$ for $k \geq 0$ by
\[
\begin{array}{llll}
t(0) \equalsdef 2, &
t(j) \equalsdef 2^{2^{2^{2^{t(j-1)}}}} \ \,
\mbox{for $j \geq 1$}, &
T_k \equalsdef \Sigma^{t(k)} \ \,
\mbox{for $k \geq 0$}, &
\mbox{and} \ 
T  \equalsdef \bigcup_{k \geq 0} T_k.
\end{array}
\]
The construction of $B$ will satisfy the following requirement:
\begin{eqnarray}
\label{requirement}
  B \seq T & \mbox{and} & \| B \cap T_k \| = 1 \ \, \mbox{for each $k
    \geq 0$}.
\end{eqnarray}
Fix an enumeration $\{M_j\}_{j \geq 1}$ of all DPOMs. For each $j \geq
1$, let $p_j$ be a fixed polynomial bounding the runtime of machine
$M_j$.  Without loss of generality, assume that our enumeration
satisfies for all~$j \geq 1$ that
\[
\sum_{i = 1}^{\lceil \log j \rceil} p_{i}(0^{t(j)}) < 2^{t(j)-1}.
\]
Note that this can indeed be assumed, w.l.o.g., by clocking the
machines with appropriately slow clocks as is standard. At stage $j$
of the construction, machines $M_{1}, M_{2}, \ldots , M_{\lceil \log j
  \rceil}$ will be {\em active\/} unless they have already been
canceled during earlier stages. Define the language
\[
L_{B} \equalsdef \{ 0^n \,|\, B \cap
{\Sigma}^{n-1}0 \neq \emptyset \}.
\]
Clearly, $L_B$ is in $\np^B$ and therefore in $\np^A$.
Let $B_{j-1}$ be the content of $B$ prior to stage $j$. Initially, let
$B_0$ be the empty set. Stage~$j > 0$ of the construction of~$B$ is as
follows.

\smallskip
{\samepage
\noindent
{\bf Stage {\boldmath $j$}.}~
\begin{description}
\item[Case 1:] For no active machine $M_i$ does
  $M_{i}^{\mbox{\protect\scriptsize QBF} \oplus B_{j-1}}(0^{t(j)})$
  accept. Choose the smallest string \mbox{$w_j \in
    {\Sigma}^{t(j)-1}0$} such that $w_j$ is not queried in the
  computation of $M_{i}^{\mbox{\protect\scriptsize QBF} \oplus
    B_{j-1}}(0^{t(j)})$ for any active machine~$M_i$. Set $B_j :=
  B_{j-1} \cup \{ w_j \}$.

\item[Case 2:]  There exists an active machine $M_i$ such that
  $M_{i}^{\mbox{\protect\scriptsize QBF} \oplus B_{j-1}}(0^{t(j)})$
  accepts. Let $\tilde{i}$ be the smallest such~$i$. Mark machine
  $M_{\tilde{i}}$ as canceled, and set $B_j := B_{j-1} \cup \{
  1^{t(j)} \}$.
\end{description}
{\bf End of Stage {\boldmath $j$}.}
}
\smallskip

Since we assume our enumeration of DPOMs to be slowed down so that the
sum of the runtimes of all machines that can be active at stage~$j$
and run on input $0^{t(j)}$ is strictly less than $2^{t(j)-1}$, the
string $w_j$, if needed, indeed exists.  In addition, our assumption
on having slowed down the enumeration of machines combined with the
widely spaced gaps between the lengths of the strings considered in
subsequent stages also guarantees that the single stages of the
construction do not interfere with each other, since for each
stage~$j$, no machine that is active at this stage can reach (and
query) any string of length $t(j+1)$, that is, no oracle extension at
later stages can effect the computations performed in stage~$j$.

Note further that Case~1 in this construction happens infinitely
often, as each Case~2 cancels a machine, but at stage~$j$ at most
$\lceil \log j \rceil$ machines have been active, so Case~2 can happen
at most $\lceil \log j \rceil$ times. Since Case~1 happens infinitely
often, it is clear that $L_{B}$ is an infinite set in~$\np^A$\@. It
remains to prove that (a)~\mbox{$\np^A = \mbox{PSPACE}^A$},
(b)~$\np^A$ is $\p^A$-immune, and (c)~$\np^A$ is not $\p^A$-bi-immune.

Statement~(a) follows immediately from the form of the oracle $A =
\mbox{QBF} \oplus B$ and the fact that QBF is PSPACE-complete.

To prove Statement~(b), note that each DPOM~$M_i$ is either canceled
eventually, or $M_i$ is never canceled. If $M_i$ is canceled, then
we have by construction that $0^{t(j)} \in L(M_{i}^{A})$ for some~$j$,
yet $0^{t(j)} \not\in L_{B}$, since \mbox{$B \cap {\Sigma}^{t(j)-1}0 =
  \emptyset$}. Thus, the language accepted by $M_i$ relative to~$A$,
$L(M_{i}^{A})$, is not a subset of~$L_B$. In the other case (i.e., if
$M_i$ never is canceled), we will argue that \mbox{$L(M_{i}^{A}) \cap
  L_B$} must be a finite set. Indeed, let $s_i$ be the first stage in
which all machines~$M_{\ell}$, with $\ell < i$, that will ever be
canceled are already canceled. Then, for no stage~$j$ with $j \geq
s_i$ will $M_{i}^{\mbox{\protect\scriptsize QBF} \oplus B_{j-1}}$
accept the input~$0^{t(j)}$, as otherwise $M_i$ would have been the
first (i.e., having the smallest number according to the enumeration)
active machine accepting $0^{t(j)}$ and would thus have been
canceled. It follows that $M_{i}^{A}$ accepts at most a finite number
(more precisely, at most~$s_i - 1$) of the elements of~$L_{B}$\@.  To
summarize, we have shown that there exists an infinite set in~$\np^A$
(namely,~$L_B$) having no infinite subset in~$\p^A$, that is, $\np^A$
is $\p^A$-immune.

Finally, let us prove Statement~(c). Suppose there exists an infinite
set $L$ in~$\np^A$. Define the function $r$ by 
\begin{eqnarray*}
r(0) \equalsdef 2^{2^{2}} & \mbox{and} &
r(j) \equalsdef 2^{2^{2^{2^{r(j-1)}}}} \ \ \mbox{for $j \geq 1$}.
\end{eqnarray*} 
Define the sets
\begin{eqnarray*}
  L_{\mbox{\protect\scriptsize in}}
  \equalsdef \{ 0^n \,|\, (\exists j
  \geq 0)\, [r(j) = n \,\wedge\, 0^n \in L] \} & \mbox{and} &
  L_{\mbox{\protect\scriptsize out}}
  \equalsdef \{ 0^n \,|\, (\exists j
  \geq 0)\, [r(j) = n \,\wedge\, 0^n \not\in L] \}.
\end{eqnarray*}
Clearly, 
$L_{\mbox{\protect\scriptsize in}} \seq L$ and
$L_{\mbox{\protect\scriptsize out}} \seq \overline{L}$, and either
$0^{r(j)} \in L$ for infinitely many $j$, or $0^{r(j)} \not\in L$ for
infinitely many $j$, or both.  Thus, either
$L_{\mbox{\protect\scriptsize in}}$ is an infinite subset of $L$, or
$L_{\mbox{\protect\scriptsize out}}$ is an infinite subset of
$\overline{L}$, or both.

Now we prove that both $L_{\mbox{\protect\scriptsize in}}$ and
$L_{\mbox{\protect\scriptsize out}}$ are in~$\p^A$\@. Recall that $A =
\mbox{QBF} \oplus B$ and that $B \seq T$ and $\| B \cap T_k \| = 1$,
since the construction of~$B$ satisfies
requirement~(\ref{requirement}) above. We now describe a
DPOM~$M_{\mbox{\protect\scriptsize in}}$ for which
$L(M_{\mbox{\protect\scriptsize in}}^{A}) =
L_{\mbox{\protect\scriptsize in}}$. On input~$x$,
$M_{\mbox{\protect\scriptsize in}}^{A}$ first checks whether~$x$ is of
the form~$0^{r(j)}$ for some~$j$. If not,
$M_{\mbox{\protect\scriptsize in}}^{A}$ rejects~$x$. Otherwise, assume
$x = 0^{r(k)}$ for some fixed $k \in \N$. Now
$M_{\mbox{\protect\scriptsize in}}^{A}$ constructs a potential query
table,~$Q$, of all strings in~$B$ that can be touched in the
computation of the $\np^A$ machine accepting~$L$. Note that all
strings in~$B$ that are smaller than $|x| = r(k)$ can be found by
brute force, since---by definition of the functions~$r$ and~$t$---they
have lengths at least double-exponentially smaller than~$|x|$. For the
same reason, no string in~$B$ of length not smaller than~$|x|$ can be
touched in the run of the $\np^A$ machine accepting~$L$, on input~$x$,
more than finitely often, and all those strings of \mbox{$B \cap
  \left( \bigcup_{j \geq k} T_j \right)$} that indeed are queried in
  this computation can thus be hard-coded into table~$Q$\@. Therefore,
  $Q$~contains all information of~$B$ that can effect the computation
  of the $\np^A$ machine for~$L$ on input~$x$.  Hence, again employing
  the PSPACE-completeness of~QBF, $M_{\mbox{\protect\scriptsize
      in}}^{A}$ can ask the QBF part of its oracle to simulate that
  computation, using table $Q$ for each query to~$B$. If this
  simulation returns the answer ``$x \in L$,'' then
  $M_{\mbox{\protect\scriptsize in}}$ accepts~$x$; otherwise,
  $M_{\mbox{\protect\scriptsize in}}$ rejects~$x$. The proof that
  \mbox{$L_{\mbox{\protect\scriptsize out}} \in \p^A$} is analogous
  and thus omitted. To summarize, we have shown that if there exists
  an infinite set $L$ in~$\np^A$, then at least one of $L$ or
  $\overline{L}$ must contain an infinite subset (specifically,
  $L_{\mbox{\protect\scriptsize in}}$ or $L_{\mbox{\protect\scriptsize
      out}}$) that is decidable in~$\p^A$, that is, $\np^A$ is not
  $\p^A$-bi-immune.  
\hfill$\Box$

\medskip

Note that the oracle~$A$ constructed in the previous proof is
recursive and is ``reasonable,'' since $\p \neq \np$ holds relative
to~$A$, due to the $\p^A$-immunity of $\np^A$\@. In addition, relative
to~$A$, the reverse of arrows~(1a) and~(1c) in Figure~\ref{fig:arrows}
fails and \mbox{$\mbox{FINITE} \neq \easyforall \neq \p$}, i.e.,
arrow~(10) in Figure~\ref{fig:arrows} is not invertible.

\begin{corollary}
\label{cor:nifty}
\quad
There exists a recursive oracle $A$ such that
\begin{enumerate}
\item \label{cor:nifty-1}
$\np^A$ is $\p^A$-immune, yet $\np^A \cap \conp^A$ is not
  $\p^A$-bi-immune,

\item \label{cor:nifty-2}
$\np^A \cap \conp^A \neq \p^A$, yet $\np^A \cap \conp^A$ is not
  $\p^A$-bi-immune,

\item \label{cor:nifty-3}
$\p^A \not\seq ( \easyforall )^A$, and

\item \label{cor:nifty-4}
$(\easyforall )^A \cap \pinf^A \neq \emptyset$.
\end{enumerate}
\end{corollary}

\noindent
{\bf Proof.}
\quad
The first two statements follow from Theorem~\ref{thm:cute}, since the
non-$\p^A$-bi-immunity of $\np^A$ clearly implies that $\np^A \cap
\conp^A$ is not $\p^A$-bi-immune. Moreover, since $\np^A =
\mbox{PSPACE}^A$ implies that $\np^A = \conp^A$, there exists a set in
$(\np^A \cap \conp^A) - \p^A$ by the $\p^A$-immunity of $\np^A$.
Furthermore, since $\np^A \cap \conp^A \neq \p^A$, the relativized
version of Theorem~\ref{thm:arrows}.\ref{thm:arrows-5} establishes the
third statement of this corollary: $\p^A \not\seq ( \easyforall )^A$.

\smallskip

To prove the fourth statement, let $A = \mbox{QBF} \oplus B$ be the
oracle constructed in the proof of Theorem~\ref{thm:cute}.  Recall
from that proof the definitions of the functions $r$ and $t$ and of
the sets $T_k$, $k \geq 0$, and~$T$.  Recall that the construction
of~$B$ ensures that the following requirement is satisfied: $B \seq T$
and $\|B \cap T_k\| = 1$ for every $k \geq 0$.  Define the set
\[
L \equalsdef \{ 0^{r(j)} \,|\, j \geq 0 \}.
\]
Clearly, $L$ is an infinite set in~P, and therefore
in~$\p^A$\@.

Let $N$ be any NPOM from our fixed enumeration of all NPOMs (see
the proof of Theorem~\ref{thm:cute}) that, with oracle $\mbox{QBF}
\oplus B$, accepts~$L$, i.e.,
\mbox{$L(N^{\mbox{\protect\scriptsize QBF} \oplus B}) = L$}. To
show that~$N$, with oracle $\mbox{QBF} \oplus B$, has always easy
certificates, we will describe a DPOM~$M$ that computes, using
oracle $\mbox{QBF} \oplus B$, an $\fp^{\mbox{\protect\scriptsize QBF}
  \oplus B}$ function~$f_N$ such that~$f_N$ prints an accepting path
of $N^{\mbox{\protect\scriptsize QBF} \oplus B}(x)$ for each~$x
\in L$.

On input~$x$ of the form $0^{r(j)}$ for some~$j$, $M$~computes by brute
force a potential query table, $Q$, of all ``short'' strings in~$B$,
i.e., \mbox{$Q = B^{< r(j)} = B^{\leq t(j)}$}, and then employs the
PSPACE-completeness of QBF to construct, bit by bit, the
lexicographically first accepting path of
$N^{\mbox{\protect\scriptsize QBF} \oplus B}(x)$, say~$p$, via prefix
search, where table $Q$ is used to answer all queries to~$B$. Since by
definition of~$r$ and~$t$, $|x| = r(j)$ is at least
double-exponentially smaller than any string of length~$> t(j)$, no
string in~$B$ of length~$> t(j)$ can be queried in the run of
$N^{\mbox{\protect\scriptsize QBF} \oplus B}(x)$ more than finitely
often, and all those strings in \mbox{$B \cap \left( \bigcup_{i >
    t(j)} T_i \right)$} that indeed are queried in this computation
can thus be hard-coded into table~$Q$\@. Therefore, $Q$ contains all
information of $B$ that can effect the computation
of~$N^{\mbox{\protect\scriptsize QBF} \oplus B}(x)$. It follows that
the path $p$ constructed 
by $M^{\mbox{\protect\scriptsize QBF} \oplus B}(x)$
indeed is a valid accepting path
of~$N^{\mbox{\protect\scriptsize QBF} \oplus B}(x)$.  $M$~outputs~$p$.
Hence, $L \in (\easyforall )^A$. This establishes our claim that
$(\easyforall )^A \cap \pinf^A \neq \emptyset$.  
\hfill$\Box$

\medskip

\begin{remark}
\label{rem:easyforallio}
\quad
In fact, it is not hard to see that, via using a Kolmogorov complexity
based oracle construction, we can even prove the following claim that
is stronger than Corollary~\ref{cor:nifty}.\ref{cor:nifty-3} above:
There exists an oracle $D$ such that \mbox{$\p^D \not\seq (
  \easyforallio )^D$}.  To be a bit more precise, in this proof, the set
\mbox{$L \equalsdef \{ 0^{t(j)} \,|\, j \geq 0 \}$} is clearly a~P set
(and thus, \mbox{$L \in \p^X$ for any~$X$}), but we can construct an
oracle set~$D$ such that there exists an NPOM $N$ with~\mbox{$L(N^D) =
  L$}, yet $N^D$ has only hard certificates almost everywhere, i.e.,
\mbox{$L \not\in ( \easyforallio )^D$}.
\end{remark}

\typeout{DO NOT FORGET TO FILL IN AND PROVE THIS REMARK! THANKS!}

Baker, Gill, and Solovay proved that there exists an oracle~$E$
relative to which \mbox{$\p^E \neq \np^E$}, yet \mbox{$\p^E = \np^E
  \cap \conp^E$}~\cite{bak-gil-sol:j:p-np-oracle}. Due to the priority
argument they apply, this proof is the most complicated of all proofs
presented in their paper.  Next, we show that an adaptation of their
proof yields the stronger result that the implication
\mbox{$(\p \neq \np \Lora \p \not\seq \easyforall)$} (which is the
reverse of arrow~(6) in Figure~\ref{fig:arrows}) fails in some
relativized world. It is worth noting that thus already the trivial part of
the implication chain \mbox{$(\p \neq \np \cap \conp \Lora \p \not\seq
  \easyforall )$} and \mbox{$(\p \not\seq \easyforall \Lora \p \neq
  \np )$} (arrows~(5) and~(6) in Figure~\ref{fig:arrows}) is shown
to be irreversible up to the limits of relativizing techniques.
Furthermore, by inserting into the obvious implication \mbox{$(\p \neq
  \np \cap \conp \Lora \p \neq \np )$} the statement ``\mbox{$\p
  \not\seq
  \easyforall$},''\footnote{%
\protect\singlespacing
We stress that it is non-trivial to show that this insertion indeed 
is possible, due to the commonly agreed non-triviality of Borodin and 
Demers's result~\cite{bor-dem:t:selfreducibility} (see 
Theorem~\ref{thm:bor-dem} and the proof of 
Theorem~\ref{thm:arrows}.\ref{thm:arrows-5}).
}
we have, on the one hand, enlightened the close connection between the
famous classic results of~\cite{bak-gil-sol:j:p-np-oracle}
and~\cite{bor-dem:t:selfreducibility}. On the other hand, having inserted
``\mbox{$\p \not\seq \easyforall$}'' into this implication 
clearly distinguishes the domains
where the hard, non-trivial cores of the proofs of the respective
results in~\cite{bak-gil-sol:j:p-np-oracle}
and~\cite{bor-dem:t:selfreducibility} fall into.

\begin{theorem}
\label{thm:bgs-like}
\quad 
There exists a recursive oracle~$A$ such that 
$\np^A \neq \p^A = ( \easyforall )^A$.
\end{theorem}

\begin{corollary}
\label{cor:bgs-like}
\cite{bak-gil-sol:j:p-np-oracle}
\quad
There exists a recursive oracle~$A$ such that $\np^A \neq \p^A = \np^A
\cap \conp^A$.
\end{corollary}

\noindent
{\bf Proof of Corollary~\ref{cor:bgs-like}.}
\quad
This follows from Theorem~\ref{thm:bgs-like} and the fact that the
proof of Theorem~\ref{thm:arrows}.\ref{thm:arrows-5} relativizes, that
is, stating the contrapositive of
Theorem~\ref{thm:arrows}.\ref{thm:arrows-5}: For every oracle~$B$, if
\mbox{$\p^B \seq ( \easyforall )^B$}, then \mbox{$\p^B = \np^B \cap
  \conp^B$}.  
\hfill$\Box$

\medskip

\noindent
{\bf Proof of Theorem~\ref{thm:bgs-like}.} 
\quad
The oracle $A$ will again be $\mbox{QBF} \oplus B$, where QBF is any
fixed PSPACE-complete problem, $B \equalsdef \bigcup_{n \geq 0}
B_n$ is constructed in stages, and for every $n > 0$, $B_{n-1}$
denotes the content of $B$ prior to stage~$n$. Initially, set $B_0$ to 
be the empty set. To make clear how to construct~$B$, we
will recall the crucial parts of Baker, Gill, and Solovay's oracle
construction in the proof
of~\mbox{\cite[Theorem~6]{bak-gil-sol:j:p-np-oracle}}, pointing out the
differences to our construction.

As in~\cite{bak-gil-sol:j:p-np-oracle}, define the function~$e$ by
\mbox{$e(0) \equalsdef 2$} and \mbox{$e(j) \equalsdef 2^{2^{e(j-1)}}$}
for~$j \geq 1$. At stage $n$ of the construction, at most one string
of length $e(n)$ is added to $B$ so as to simultaneously ensure both
$\p^A \neq \np^A$ and $\p^A \seq ( \easyforall )^A$.

Fix an enumeration $\{M_j\}_{j \geq 1}$ of all DPOMs and an
enumeration $\{N_j\}_{j \geq 1}$ of all NPOMs. For each~$j \geq 1$,
let $p_j$ be a fixed polynomial bounding the runtime of both $M_j$
and~$N_j$. 
Define the language
\[
L_{B} \equalsdef \{ 0^n \,|\, (\exists w)\, [w \in B^{=n}] \}.
\]
Clearly, $L_B$ is in $\np^B$, and therefore in $\np^A$\@.  

There are two types of requirements to be satisfied in the
construction of~$B$. Intuitively, satisfying a requirement of the form
$\< i, i \>$ will ensure that \mbox{$L(M_{i}^{A}) \neq L_B$}. Thus,
satisfying $\< i, i \>$ for each $i \geq 1$ will establish our claim
that \mbox{$\p^A \neq \np^A$}\@. On the other hand, satisfying a
requirement of the form $\< j, k \>$ with $j \neq k$ will ensure that
\mbox{$L(N_{k}^{A}) \neq L(M_{j}^{A})$}, and $N_{k}^{A}$ is thus not a
machine accepting the $\p^A$ set $L(M_{j}^{A})$\@.  Therefore, showing
that $N_{k}^{A}$ has always easy certificates for every requirement
$\< j, k \>$ with $j \neq k$ that is {\em never\/} satisfied (and thus
the equality \mbox{$L(N_{k}^{A}) = L(M_{j}^{A})$} might happen) will
suffice to establish our claim that \mbox{$\p^A \seq ( \easyforall
  )^A$}\@.

In more detail, an unsatisfied requirement $\< i, i \>$ is {\em
  vulnerable\/} at stage~$n$ of the construction of~$B$ if
\mbox{$p_{i}(e(n)) < 2^{e(n)}$}. An unsatisfied requirement $\< j, k
\>$ with $j \neq k$ is {\em vulnerable\/} at stage~$n$ if there exists
a string~$x$ such that
\begin{eqnarray}
\label{equ:vulnerable}
e(n-1) < \log |x| \leq e(n) \leq \max\{ p_{j}(|x|), p_{k}(|x|) \} <
e(n+1)
\end{eqnarray}
and in addition it holds that 
\mbox{$x \in L(M_{j}^{\mbox{\protect\scriptsize QBF} \oplus B_{n-1}})$}
if and only if \mbox{$x \not\in L(N_{k}^{\mbox{\protect\scriptsize
      QBF} \oplus B_{n-1}})$}.  We note that the definition of
vulnerability for this second type of an unsatisfied requirement is
different from that given in the proof
of~\cite[Theorem~6]{bak-gil-sol:j:p-np-oracle}. By convention, we
agree that requirement~$R_1$ has higher priority than
requirement~$R_2$ exactly if~$R_1 < R_2$.  Stage~$n > 0$ of the
construction of~$B$ is as follows.

\smallskip
{\samepage
\noindent
{\bf Stage {\boldmath $n$}.} \quad
The requirement of highest priority that is vulnerable at stage~$n$
will be satisfied. To satisfy requirement $\< j, k \>$ with $j \neq
k$, 
we simply add no string to~$B$ in this stage, i.e., $B_n := B_{n-1}$. 
To satisfy requirement
$\< i, i \>$, simulate the computation of
$M_{i}^{\mbox{\protect\scriptsize QBF} \oplus B_{n-1}}(0^{e(n)})$. If
it rejects, then let $w_n$ be the smallest string of length $e(n)$
that is not queried along the computation of
$M_{i}^{\mbox{\protect\scriptsize QBF} \oplus B_{n-1}}(0^{e(n)})$, and
set \mbox{$B_n := B_{n-1} \cup \{w_n\}$}.  If
$M_{i}^{\mbox{\protect\scriptsize QBF} \oplus B_{n-1}}$ accepts
$0^{e(n)}$, then set $B_n := B_{n-1}$.\\ 
{\bf End of Stage {\boldmath $n$}.} 
} %
\smallskip

Each requirement $\< i, i \>$ is eventually satisfied, since there are
only finitely many requirements of higher priority. Suppose
requirement $\< i, i \>$ is satisfied at stage~$n$. Then, since $\< i,
i \>$ is vulnerable at stage~$n$, we have \mbox{$p_{i}(e(n)) <
  2^{e(n)}$}. This implies that the string~$w_n$, if needed to be
added to~$B$ in stage~$n$, must exist, and further that no string
in~$B$ of length~$>e(n)$ can be touched in the run of
$M_{i}^{A}(0^{e(n)})$. Since by construction also~$w_n$ is not 
queried by $M_{i}^{A}(0^{e(n)})$, we conclude that oracle extensions
at stages~$\geq n$ do not effect the computation of
$M_{i}^{\mbox{\protect\scriptsize QBF} \oplus B_{n-1}}(0^{e(n)})$.
Hence, \mbox{$0^{e(n)} \not\in L(M_{i}^{\mbox{\protect\scriptsize QBF}
    \oplus B})$} if and only if \mbox{$0^{e(n)} \not\in
  L(M_{i}^{\mbox{\protect\scriptsize QBF} \oplus B_{n-1}})$} if and
only if there exists some string $w_n \in B^{=e(n)}$ if and only if
$0^{e(n)} \in L_B$; so \mbox{$L(M_{i}^{\mbox{\protect\scriptsize QBF}
    \oplus B}) \neq L_B$}.  It follows that $L_B \in \np^A - \p^A$.

It remains to prove that \mbox{$\p^A \seq ( \easyforall )^A$}. The
remainder of this proof is different from the proof
in~\cite{bak-gil-sol:j:p-np-oracle}.  Given any pair of machines,
$M_j$ and $N_k$ with~$j \neq k$, we will show that either
\mbox{$L(N_{k}^{A}) \neq L(M_{j}^{A})$}, or $N_{k}^{A}$ has always
easy certificates, thus proving that for every set in $\p^A$, each
$\np^A$ machine accepting it has always easy certificates, in symbols,
\mbox{$\p^A \seq ( \easyforall )^A$}.

Clearly, each requirement $\< j, k \>$ with $j \neq k$ is either
satisfied eventually, or $\< j, k \>$ is never satisfied.  If
requirement $\< j, k \>$ is satisfied at stage~$n$ for some~$n$, 
then we are done,
since there exists a string~$x$ in this case such that 
(i)~\mbox{$x \in L(M_{j}^{\mbox{\protect\scriptsize QBF} \oplus B_{n-1}})$} 
    if and only if \mbox{$x \not\in L(N_{k}^{\mbox{\protect\scriptsize QBF}
    \oplus B_{n-1}})$},
(ii)~$B_n = B_{n-1}$, and
(iii)~neither $M_{j}$ nor $N_{k}$ can query any string of length 
$> e(n)$ on input~$x$.
Thus, \mbox{$x \in
  L(M_{j}^{A})$} if and only if \mbox{$x \not\in L(N_{k}^{A})$}, i.e.,
$N_{k}^{A}$ cannot accept the $\p^A$ set~$L(M_{j}^{A})$. So
suppose requirement $\< j, k \>$ is never satisfied, i.e.,
\mbox{$L(N_{k}^{A}) = L(M_{j}^{A})$} might now happen. Then,
it suffices to show that \mbox{$L(N_{k}^{A}) = L(M_{j}^{A})$} implies
that $N_{k}^{A}$ has always easy certificates.  Since this holds for
all~$k$ for which $N_{k}^{A}$ can accept $L(M_{j}^{A})$, 
we have \mbox{$L(M_{j}^{A}) \in ( \easyforall )^A$}.

Let $s_{j,k}$ be the first stage such that
\begin{description}
\item[(a)] for every $x$ such that $|x| \geq e(s_{j,k})$ there is at
  most one~$n$ such that
\[
\log |x| \leq e(n) \leq \max\{ p_{j}(|x|), p_{k}(|x|) \},
\]

\item[(b)] all requirements of higher priority than $\< j, k \>$ that
  will ever be satisfied are already satisfied.
\end{description}

We will now show that $N_{k}^{A}$ on input~$x$ has easy certificates
for every string~$x$ accepted by~$M_{j}^{A}$\@. We describe an $\fp^A$
function $f_k$ that, on input~$x$, uses oracle $A = \mbox{QBF} \oplus B$ 
to output some
accepting path of~$N_{k}^{A}(x)$ if $x \in L(M_{j}^{A})$\@.

On input~$x$, if $|x| < e(s_{j,k})$, then $f_k$ uses a finite table to
find and output some accepting path of $N_{k}^{A}(x)$ whenever $x \in
L(M_{j}^{A})$\@. Otherwise (i.e., if~$|x| \geq e(s_{j,k})$), $f_k$
calculates the smallest~$n$ such that $e(n) \geq \log |x|$. Then,
$f_k$ builds a table, $T$, of all strings that were added to $B$
before stage~$n$, i.e., \mbox{$T = B^{< e(n)}$}, by querying its
oracle~$B$ about {\em all\/} strings of lengths $e(0), e(1), \ldots ,
e(n-1)$.  Since $e(n-1) < \log |x|$, only ${\cal O}(|x|)$ queries are
required in this brute-force search. We have to consider two cases.

\smallskip
\begin{description}
\item[Case 1:] $e(n) > \max\{ p_{j}(|x|), p_{k}(|x|) \}$.  Then,
  neither $M_{j}^{A}(x)$ nor $N_{k}^{A}(x)$ can query their oracle
  about any string of length~$\geq e(n)$. Therefore, the computation
  of $M_{j}$ and $N_{k}$ on input~$x$ with oracle \mbox{$\mbox{QBF}
    \oplus T$} is the same as with oracle \mbox{$\mbox{QBF} \oplus
    B$}. Hence, $f_k$ can run $M_{j}^{\mbox{\protect\scriptsize QBF}
    \oplus T}$ on input~$x$ to determine whether $M_{j}^{A}$ would
  accept $x$. If it rejects $x$, then $f_k$ can output an arbitrary
  string, and we are done. If $M_{j}^{\mbox{\protect\scriptsize QBF}
    \oplus T}$ accepts~$x$, then $f_k$ exploits the PSPACE-completeness
  of QBF to construct the lexicographically first accepting path of
  $N_{k}^{\mbox{\protect\scriptsize QBF} \oplus B}(x)$, say~$p$, bit
  by bit via prefix search, where QBF uses table $T$ to answer every
  oracle call of $N_{k}^{\mbox{\protect\scriptsize QBF} \oplus B}(x)$
  to~$B$.  It follows that~$p$ is a valid accepting path of $N_{k}^{A}(x)$
    if~$x \in L(M_{j}^{A})$. $f_k$~outputs~$p$.

\item[Case 2:] $e(n) \leq \max\{ p_{j}(|x|), p_{k}(|x|) \}$.  In this
  case, also strings of length~$e(n)$ can be queried by $M_{j}^{A}(x)$
  or $N_{k}^{A}(x)$. Clearly, $N_{k}^{\mbox{\protect\scriptsize
        QBF} \oplus T}$ accepts~$x$ if and only if 
  $M_{j}^{\mbox{\protect\scriptsize QBF} \oplus T}$ accepts~$x$, as 
  otherwise requirement $\< j, k \>$ would have
  been satisfied at stage~$n$, contradicting our supposition that 
  $\< j, k \>$ is never satisfied.  Indeed, since $\< j, k \>$ is the
  smallest unsatisfied requirement at stage~$n$ by~(b) above
  and since~$x$ meets condition~(\ref{equ:vulnerable}) above by~(a),
  the equivalence
  $$\left(x \in L(M_{j}^{\mbox{\protect\scriptsize QBF} \oplus T}) 
  \Lolra x \not\in L(N_{k}^{\mbox{\protect\scriptsize QBF} \oplus T})\right)$$
  would enforce the vulnerability of $\< j, k \>$ at this stage. Now,
$f_k$ simulates $M_{j}^{A}(x)$ and outputs an arbitrary string if it
rejects. Otherwise (i.e., if \mbox{$x \in L(M_{j}^{A})$}), 
$f_k$ runs $M_{j}^{\mbox{\protect\scriptsize QBF} \oplus T}(x)$, call
this computation~$q$. There are two subcases.

\begin{description}
\item[Case 2.1:] The computation of $M_{j}^{A}(x)$ exactly agrees with~$q$.
  Then, there exists an accepting path of
  $N_{k}^{\mbox{\protect\scriptsize QBF} \oplus T}(x)$, and 
  $f_k$ again employs QBF to construct the
  lexicographically first accepting path of
  $N_{k}^{\mbox{\protect\scriptsize QBF} \oplus T}(x)$, call this
  path~$p$. If~$p$ were reliable w.r.t.\ oracle~$A$, then
  $f_k$ could simply output~$p$, and we were done. 
  However, $p$ is not reliable, since~$T$ and~$B$
  might differ, so~$f_k$ has to check the validity of~$p$.  By
  our choice of $s_{j,k}$, there exists (according to (a) above) at
  most one~$n$ such that \mbox{$\log |x| \leq e(n) \leq \max\{
    p_{j}(|x|), p_{k}(|x|) \}$}.  Hence, $T$ can lack at most one
  length~$e(n)$ string of~$B$ that might be queried in the run of
  $N_{k}^{A}(x)$. Now, $f_k$~checks whether~$p$ is a valid
  certificate of~$N_{k}^{A}(x)$ by simply checking whether all answers
  given along~$p$ are correct according to the~$B$ part of~$f_k$'s
  oracle. There are two subcases of Case~2.1.

\begin{description}
\item[Case 2.1.1:] All strings $z$ queried along~$p$ receive the
  answer ``yes'' if and only if~$z \in B$. We conclude that~$p$ 
  is a valid certificate of~$N_{k}^{A}(x)$. $f_k$~outputs~$p$.

\item[Case 2.1.2:] There exists a string~$z$ that is queried along~$p$, 
  but receives a wrong ``no'' answer according to~$B$,
  i.e.,~$z \in B$\@. Then, $f_k$ has detected that~$z$ is the one string of
  length $e(n)$ in~$B - T$. So, adding~$z$ to~$T$, we now have
  \mbox{$T = B^{\leq e(n)}$}. $f_k$~again employs 
  QBF to construct the lexicographically first
  accepting path of~$N_{k}^{\mbox{\protect\scriptsize QBF} \oplus
    T}(x)$, say~$p'$, which now must be a valid certificate of~$N_{k}^{A}(x)$. 
  $f_k$~outputs~$p'$.
\end{description}

\item[Case 2.2:] The computation of $M_{j}^{A}(x)$ differs from~$q$.
  The only way this could happen, however, is that the one missing
  string in~$T$, \mbox{$z \in B^{\leq e(n)} - T$}, is queried on~$q$, but
  has received a wrong ``no'' answer from~$T$.  Then, as in
  Case~2.1.2, $f_k$~has identified~$z$ and can complete table~$T$ by
  adding~$z$ to~$T$\@. Now, \mbox{$T = B^{\leq e(n)}$}, and $f_k$ can
  proceed as in Case~2.1.2 to find and output a valid certificate
  of~$N_{k}^{A}(x)$ if~$x$ is accepted by~$M_{j}^{A}$ (via 
  once more employing QBF in the prefix
  search to construct the lexicographically first certificate).
\end{description}
\end{description}
\smallskip

Since \mbox{$\max\{ p_{j}(|x|), p_{k}(|x|) \} < e(n+1)$} by~(a) above,
no string of length~\mbox{$\geq e(n+1)$} can be queried by~$N_{k}$
or~$M_{j}$ on input~$x$, and thus oracle extensions at
stages~\mbox{$\geq n+1$} cannot effect the computation
of~$N_{k}^{\mbox{\protect\scriptsize QBF} \oplus B_{n}}(x)$
or~$M_{j}^{\mbox{\protect\scriptsize QBF} \oplus B_{n}}(x)$.  This
completes the proof.  
\hfill$\Box$

\begin{remark}
\label{rem:subtle}
\begin{enumerate}
\item The argument given in the above proof could almost seem to even
  prove that there exists some oracle $A$ relative to which $\p^A \neq
  \np^A$ and $\np^A \seq ( \easyforall )^A$, i.e., \mbox{$\p^A \neq
    \np^A$} and \mbox{$\p^A = \np^A$}, which clearly is impossible. In
  fact, our argument would {\em not\/} work for NP in place of P, for
  the following subtle reason. When we define the vulnerability of
  requirements of the form $\< j, k\>$ with $j \neq k$ in terms of
  pairs of two {\em nondeterministic\/} oracle machines~$N_j$
  and~$N_k$, and modify our argument appropriately, then the~$\fp^A$
  function~$f_k$ has no way of telling whether or not the input~$x$ is
  accepted by $N_{j}^{A}$ (as we have no $\p^A$ algorithm as in the
  above proof) and therefore is in serious trouble when it is trying
  to construct a valid certificate of~$N_{k}^{A}(x)$.

\item Fortnow and Rogers~\cite{for-rog:c:separability}
  have presented an oracle $A$ such that the
  reverse of our arrow~(5) in Figure~\ref{fig:arrows} fails relative
  to~$A$, and in fact they show that \mbox{$\p \seq \easyforall$} holds
  relative to any ``sparse generic oracle with the {\em subset\/}
  property.''  In fact, regarding their oracle~$A$,
  Fenner et al.~\cite{fen-for-nai-rog:c:inverse} note that
  since they even prove that an intermediate condition (of our
  arrow~(5)) cannot imply \mbox{$\p \neq \np \cap \conp$}, they have a
  statement slightly stronger than that the reverse of arrow~(5) fails
  relative to~$A$.
\end{enumerate}
\end{remark}

Of course, the existence of relativized worlds $A$ in which a
statement $X^A$ fails should not be viewed as evidence that $X$ fails
in the unrelativized world.  Rather, the existence of such relativized
worlds should be viewed as evidence that most standard proof
techniques lack the power to prove that $X$ holds in the unrelativized
world (see, e.g.,
\cite{all:c:oracles,for:j:relativization,har:j:conflicting-relativizations}
for discussions of how to interpret relativized results).  We suggest
as an open question the issue of whether even stronger implications
than those of Figure~\ref{fig:arrows} can be established.

As a final remark, an anonymous referee mentioned the interesting open
topic of definitions analogous to ours, except with the path-finding
operator being probabilistic (either error-bounded or
error-unbounded), rather than deterministic.

{\samepage
\begin{center}
{\bf Acknowledgments}
\end{center}
\nopagebreak
\noindent
We thank Eric Allender for pointing out that if $\np \cap \conp$ is
P-bi-immune, then \mbox{$\easyforall = \mbox{FINITE}$}, for pointing
out an alternative (and in fact stronger) proof of 
Theorem~\ref{thm:arrows}.\ref{thm:arrows-4}, and for
generally inspiring this line of research. We acknowledge interesting
discussions with Alan Selman and Lance Fortnow on this subject. In
particular, we are indebted to Alan Selman for generously permitting
us to include his proof of Claim~\ref{sel:perscomm} and to Lance
Fortnow for providing us with an advance copy of the
manuscript~\cite{fen-for-nai-rog:c:inverse}.

}%

{\singlespacing

{\bibliography{main.bbl}}
}

\end{document}